\documentclass[12pt]{iopart}
\usepackage{setstack}
\usepackage{amssymb}
\usepackage{graphicx}
\begin{document}

\def\PRL#1{{\it Phys.\ Rev.\ Lett.}\ {\bf#1}}
\def\JCAP#1#2{#2 {\it J.\ Cosmol.\ Astropart.\ Phys.}\ JCAP\,{#1}\,(#2)\,}
\def\ApJ#1{{\it Astrophys.\ J.}\ {\bf#1}}
\def\PL#1#2{{\it Phys.\ Lett.\ #1}~{\bf#2}}
\def\PR#1#2{{\it Phys.\ Rev.}\ #1~{\bf#2}}\def\MPL#1#2{{\it Mod}.\ \PL{#1}{#2}}
\def\AaA#1{{\it Astron.\ Astrophys.}\ {\bf#1}}
\def\MNRAS#1{{\it Mon.\ Not.\ R.\ Astr.\ Soc.}\ {\bf#1}}
\def\CQG#1{{\it Class.\ Quantum Grav.}\ {\bf#1}}
\def\GRG#1{{\it Gen.\ Relativ.\ Grav.}\ {\bf#1}}
\def\IJMP#1#2{{\it Int.\ J.\ Mod.\ Phys.}\ #1~{\bf#2}}
\def\JMP#1{{\it J.\ Math.\ Phys.}\ {\bf #1}}
\def\RMP#1{{\it Rev.\ Mod.\ Phys.}\ {\bf#1}}
\def\beq{\begin{equation}} \def\eeq{\end{equation}}
\def\bea{\begin{eqnarray}} \def\eea{\end{eqnarray}}
\font\bm=cmmib10 \def\B#1{\hbox{\bm#1}} \def\Bom{\B{\char33}}
\def\Z#1{_{\lower2pt\hbox{$\scriptstyle#1$}}} \def\w#1{\,\hbox{#1}}
\def\X#1{_{\lower2pt\hbox{$\scriptscriptstyle#1$}}} 
\def\metric#1{g^{\raise1pt\hbox{\sevenrm #1}}_{\mu\nu}}
\font\sevenrm=cmr7 \def\ns#1{_{\hbox{\sevenrm #1}}} \def\dOM{\dd\Omega^2}
\def\Ns#1{\Z{\hbox{\sevenrm #1}}} \def\ave#1{\langle{#1}\rangle}
\def\lsim{\mathop{\hbox{${\lower3.8pt\hbox{$<$}}\atop{\raise0.2pt\hbox{$\sim$}}
$}}} \def\gsim{\mathop{\hbox{${\lower3.8pt\hbox{$>$}}\atop{\raise0.2pt\hbox{$
\sim$}}$}}} \def\kms{\w{km}\;\w{sec}^{-1}}\def\kmsMpc{\kms\w{Mpc}^{-1}}
\def\dd{{\rm d}} \def\ds{\dd s} \def\etal{{\em et al}}\def\ta{\tau}
\def\al{\alpha}\def\be{\beta}\def\ga{\gamma}\def\de{\delta}
\def\et{\eta}\def\th{\theta}\def\ph{\phi}\def\rh{\rho}\def\si{\sigma}
\def\frn#1#2{{\textstyle{#1\over#2}}} \def\Deriv#1#2#3{{#1#3\over#1#2}}
\def\Der#1#2{{#1\hphantom{#2}\over#1#2}} \def\pt{\partial} \def\ab{{\bar a}}
\def\goesas{\mathop{\sim}\limits} \def\tv{\ta\ns{v}} \def\tw{\ta\ns{w}}
\def\gb{\bar\ga} \def\I{{\hbox{$\scriptscriptstyle I$}}}
\def\av{{a\ns{v}\hskip-2pt}} \def\aw{{a\ns{w}\hskip-2.4pt}}\def\Vav{{\cal V}}
\def\DD{{\cal D}}\def\gd{{{}^3\!g}}\def\half{\frn12}\def\Rav{\ave{\cal R}}
\def\QQ{{\cal Q}} \def\rw{r\ns w}
\def\mean#1{{\vphantom{\tilde#1}\bar#1}} \def\bx{{\mathbf x}}
\def\bH{\mean H}\def\Hb{\bH\Z{\!0}} \def\gb{\mean\ga} \def\bq{\mean q}
\def\rhb{\mean\rh}\def\OM{\mean\Omega}\def\etb{\mean\eta}
\def\fw{{f\ns w}}\def\fv{{f\ns v}} \def\goesas{\mathop{\sim}\limits}
\def\fvn{f\ns{v0}} \def\fvf{\left(1-\fv\right)} \def\Hh{H}
\def\OMM{\OM\Z M}\def\OMk{\OM\Z k}\def\OMQ{\OM\Z{\QQ}}
\def\la{\lambda}\def\dL{d\Z L} \def\Hm{H\Z0}
\def\etw{\eta\ns w} \def\etv{\eta\ns v}
\def\fvi{{f\ns{vi}}} \def\fwi{{f\ns{wi}}} \def\gw{\gb\ns w}\def\gv{\gb\ns v}
\def\Hv{H\ns v} \def\Hw{H\ns w} \def\kv{k\ns v}
\def\LCDM{$\Lambda$CDM} \def\Aa{{\cal A}} \def\OmMn{\Omega\Z{M0}}
\def\GA{\Gamma}\def\SI{\Sigma}\def\Az{{\cal A}}\def\Tz{{\mathbf T}}
\def\Vv{{\cal V}}\def\Mm{{\cal M}}\def\K{{\cal K}}\def\NN{{\cal N}}
\def\figfi{\centerline{\scalebox{0.75}{\includegraphics{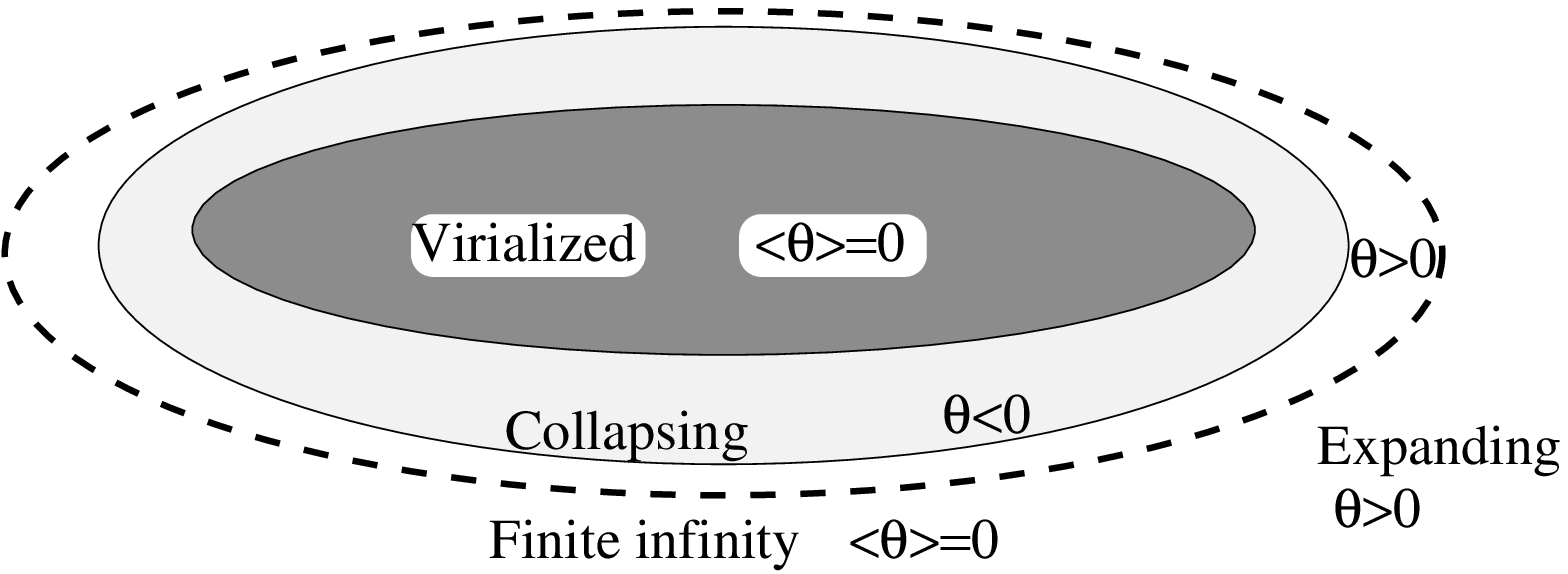}}}}
\def\figaccel{\centerline{\scalebox{0.5}{\includegraphics{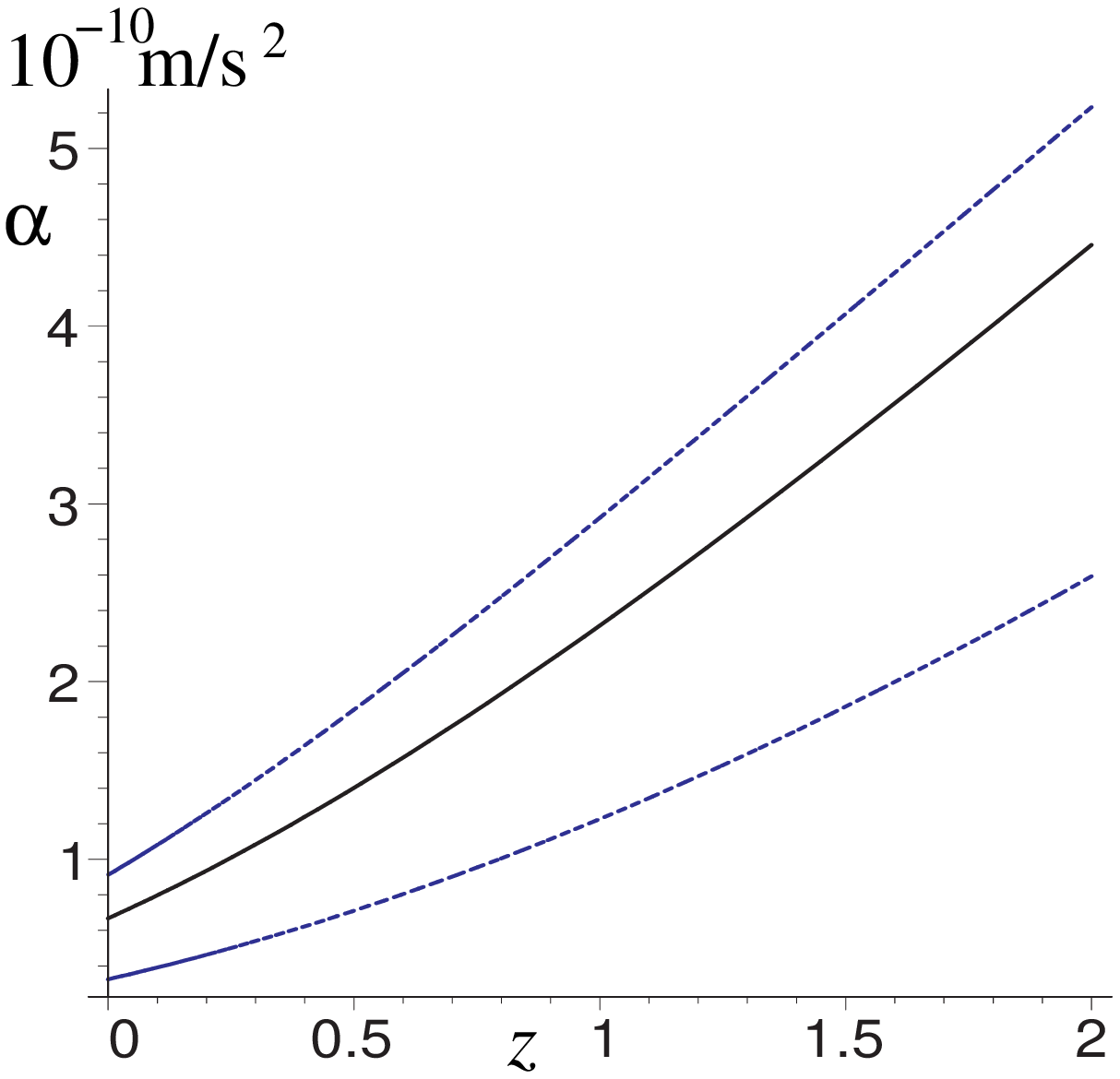}}
\hbox to1truecm{\hfil}
\scalebox{0.5}{\includegraphics{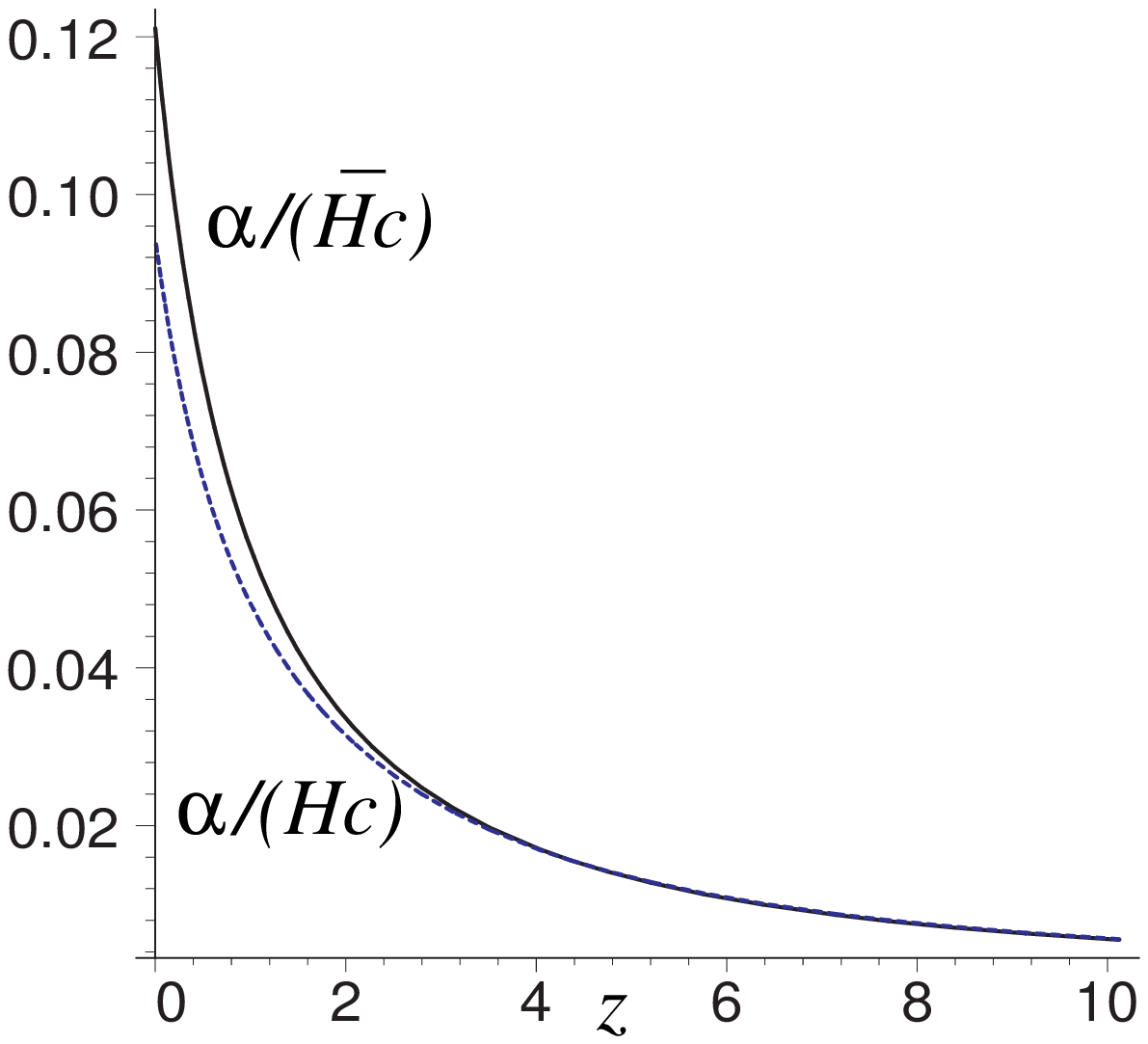}}}}
\def\Fi{\mathop{\hbox{\footnotesize\it fi}}}
\review[What is dust? -- Physical foundations of the averaging problem in
cosmology]{What is dust? -- Physical foundations of the averaging problem in
cosmology}
\author{David L. Wiltshire}
\address{Department of Physics and Astronomy, University of Canterbury,
Private Bag 4800, Christchurch 8140, New Zealand}
\eads{\mailto{David.Wiltshire@canterbury.ac.nz}\\
http://www2.phys.canterbury.ac.nz/$\goesas$dlw24/}

\begin{abstract}
The problems of coarse-graining and averaging of inhomogeneous cosmologies,
and their backreaction on average cosmic evolution, are reviewed from a
physical viewpoint. A particular focus is placed on comparing different notions
of average spatial homogeneity, and on the interpretation of observational
results. Among the physical questions we consider are: the nature of an average
Copernican principle, the role of Mach's principle, the issue of quasilocal
gravitational energy and the different roles of spacetime, spatial and null
cone averages. The observational interpretation of the timescape scenario is
compared to other approaches to cosmological averaging, and outstanding
questions are discussed.
\end{abstract}
\pacs{98.80.-k, 04.20.Cv, 98.80.Jk, 98.80.Es}\bigskip

\noindent Invited review accepted for publication in the {\em Classical and
Quantum Gravity} special issue ``Inhomogeneous Cosmological Models and
Averaging in Cosmology''.
\maketitle

\section{Introduction}

Observations show that although the universe was remarkably homogeneous at
the epoch of last scattering, when the cosmic microwave background (CMB)
radiation was laid down, at the present epoch the matter distribution displays
a very complex structure with significant inhomogeneities up to scales of at
least $100h^{-1}$ Mpc, where $h$ is the dimensionless parameter related to the
Hubble constant by $\Hm=100h\kmsMpc$. The present universe is dominated in
volume by voids \cite{HV1,HV2,Pan11}, with galaxy clusters grouped in sheets
and filaments that surround the voids, and thread them. At the largest of
scales we see a few peculiar structures, such as the Sloan Great Wall.

At the same time, despite a number of nagging puzzles, most of the gross
features of the universe are extremely well described by a spatially
homogeneous and isotropic Friedmann--Lema\^{\i}tre--Robertson--Walker (FLRW)
model, with additional Newtonian perturbations evolved by $N$--body computer
simulations to model the structure. The price that is paid for observational
concordance is that most of the matter content in the universe must be in forms
that have never been directly observed: 20--25\% in the form of clumped
nonbaryonic dark matter, and 70--75\% in the form of a smooth dark energy, with
an equation of state, $P=w\rh c^2$, extremely close to that of a cosmological
constant, $w=-1$.

The dichotomy that the universe displays considerable inhomogeneity, while
still being phenomenologically well fit by an average spatially homogeneous
evolution, has led to considerable interest in the averaging problem in
inhomogeneous cosmology. Is it possible that one or more of the components
of the dark stuff introduced for the purposes of a phenomenological fit to
observations are simply an artefact of us misunderstanding the workings
of gravity on the largest scales? Whereas some researchers immediately
leap to the extreme of modifying the whole theory of gravity, those more
intimately acquainted with general relativity are aware that the
implementation of the physical ingredients of Einstein's theory has not been
precisely specified on all scales. There are many unsolved problems provided
by the questions of coarse-graining, fitting, averaging and the
statistical notions of gravitational energy and entropy, which must inevitably
enter when dealing with the complex many--body problem that observational
cosmology presents us. These are hard problems. However, we should try to
understand the universe we observe rather than inventing toy models
purely because they are simple to solve.

It is my view that future progress in the averaging problem demands advances
in conceptual understanding. This paper will therefore review the present state
of play in averaging with a strong conceptual bias, focusing on questions
rather than answers. It is not a review of the details of mathematical
techniques in averaging; there are already a number of recent reviews of that
nature -- including, for example, those of Buchert \cite{Brev,Brev2} and van
den Hoogen \cite{vdH}. The possible mathematical choices one can make are many,
but they each entail physical choices, either explicitly or implicitly. It is
the nature of these choices that I wish to focus on.

\section{The fitting problem: On what scale are Einstein's equations
valid?}

Einstein's field equations
\beq
{G^\mu}_\nu={8\pi G\over c^4}{T^\mu}_\nu
\label{Efe}
\eeq
define the structure of general relativity as a relationship between geometry
and matter. However, the scale over which matter fields are coarse--grained
to produce the energy--momentum tensor on the r.h.s.\ of
(\ref{Efe}) is not prescribed, leaving an inherent ambiguity in the theory.
Observation provides no direct guide in this matter, since general relativity
is only well tested for isolated systems -- such as the solar system or
binary pulsars -- for which ${T^\mu}_\nu=0$. Indeed, Wheeler's aphorism that
{\em``matter tells space how to curve''} is really only tested to the extent
that matter is defined by boundary conditions and symmetry assumptions as long
as the vacuum Einstein equations apply\footnote{In the Schwarzschild geometry,
for example, a stellar interior solution is assumed to be matched with
junction conditions at the spherical surface of the star. The nature of the
stellar interior is irrelevant for the exterior vacuum geometry, however, on
account of Birkhoff's theorem.}.

Einstein's equations are designed to reduce to Poisson's equation
\beq
\nabla^2\Phi=4\pi G\rh
\label{Poisson}
\eeq
in the Newtonian limit that the spacetime geometry is that of a weak field
near a flat Minkowski background, and all characteristic velocities are
much smaller than that of light, so that $g\Z{00}=1-2\Phi/c^2$, with $\Phi\ll
c^2$, and $\rh c^2\equiv T^{00}\gg|T^{0i}|\gg|T^{ij}|$, where Latin indices
denote spatial components.

There is no ambiguity in applying the full Einstein equations (\ref{Efe})
to a fluid of particles with well-defined properties, such as ions, atoms
and molecules in the early phases of the universe's expansion. However, as
soon as gravitational collapse occurs then the geodesics of atoms and molecules
cross. There are phase transitions, and the definition of the particles in the
fluid approximation must change, giving rise to at least the following
layers of coarse-graining in the epochs following last scattering:
\begin{enumerate}
\item Atomic, molecular, ionic or nuclear particles: applicable with
\begin{itemize}
\item dust equation of state within any expanding regions which have not yet
undergone gravitational collapse;
\item fluid equation of state within relevant collapsed objects (stars, white
dwarfs, neutron stars) for periods of time between phase transitions that
alter the nongravitational particle interactions and the equation of state;
\end{itemize}
\item Collapsed objects such as stars and black holes coarse-grained as
isolated objects;
\item Stellar systems coarse-grained as dust particles within galaxies;
\item Galaxies coarse-grained as dust particles within clusters;
\item Clusters of galaxies coarse-grained as bound systems within expanding
walls and filaments;
\item Voids, walls and filaments combined as as expanding regions of different
densities in a single smoothed out cosmological fluid.
\end{enumerate}

General relativity with the vacuum Einstein equations is well--tested at level
(ii), and it generally accepted that the Einstein equations with a microscopic
fluid $T^{\mu\nu}$ apply at level (i), with the small caveats that the
equation of state of objects with the density of neutron stars is not
completely understood, and also that general relativity must break down near
singularities in the extreme strong field regime. However, once we proceed to
higher levels in this fitting problem \cite{fit1,fit2} the physical issues
become more and more murky. Provided that we can ignore any galactic magnetic
fields etc, and only consider the effects of gravity, then at levels
(iii)--(vi) we are generally only dealing with dust sources.
However, as the definition of dust becomes less and less clear with successive
coarse-grainings, the scale on which the Einstein equations should apply
becomes open to question.

\subsection{Coarse--graining\label{grain}}
One outstanding problem is that the mathematical problem of coarse-graining
in general relativity is very little studied. Any coarse-graining procedure
amounts to replacing the the microphysics of a given spacetime region by
some collective degrees of freedom of those regions which are sufficient to
describe physics on scales larger than the coarse-graining scale. Einstein's
equations were originally formulated with the intent that the energy-momentum
tensor on the r.h.s.\ of (\ref{Efe}) should either describe fundamental fields,
such as the Maxwell field, or alternatively to the coarse-graining of the
purely nongravitational interactions described by such fields in the fluid
approximation.

Einstein originally imagined a universe with the density of the Milky Way; the
complex hierarchy of galaxies, galaxy clusters, filaments, walls and voids was
unknown when he wrote down his equations (\ref{Efe}). The fundamental problem
then, is that since the universe is composed of a hierarchy of long-lived
structures much larger than those of stars, we must also coarse-grain over
gravitational interactions within that hierarchy to arrive at a fluid
description for cosmology. With such a coarse-graining, geometry no longer
enters purely on the left hand side of Einstein's equations but in a
coarse-grained sense can be hidden inside effective fluid elements on the right
hand side.

The most fundamental quantities of interest as the sources of the right hand
side of Einstein's equations are those of mass--energy, momentum and angular
momentum. Effectively, if we demand that equations (\ref{Efe}) should
also apply in a coarse-grained version on cosmological scales, then it means
that we are seeking collective parameters such as mass--energy which
average over local spatial curvature, rotational kinetic energy etc.
Furthermore we must approach the problem more than just once, on a
succession of scales. This necessarily involves the issue of quasilocal
gravitational energy, and more particularly statistical properties of
the gravitational interactions of bound systems.

Since we are no longer dealing with a fixed spatial metric this problem is
far more complicated than it is in Newtonian theory, and indeed it is largely
unexplored territory. Before surveying what has been done, let me
outline the challenges presented to us by observation at each of the levels
in the hierarchy presented above.

In going from level (i) to level (ii), the simple coarse-graining problem for
nongravitational interactions can be ignored for stellar system
astrophysics, since by symmetry assumptions in a vacuum spacetime we can
solve the Einstein equations exactly, and leave the matter to be assumed to
be an interior solution with a fluid equation of state matched at the timelike
boundary which defines the surface of the star.

To the best of my knowledge the formal coarse-graining of vacuum geometries as
dust to proceed from level (ii) to level (iii) is not a problem that has been
directly studied. However, quasilocal local mass definitions are very well
developed for stationary asymptotically flat systems, and provided we are
sufficiently far from any isolated source then gravity is generally assumed to
coincide with its Newtonian limit via (\ref{Poisson}). Thus one can easily
envisage a smoothing procedure in which one excises a timelike worldtube with
$S^2$ spatial topology around an isolated source, and replaces it by a density
in terms of an ADM-like mass divided by the excised spatial volume. The
coarse-graining procedure of Korzy\'nski \cite{Ko09} provides a formalism in
which this might be realised.

\subsubsection{Galactic dynamics}
Already at the level (iii) of galactic dynamics general relativity offers
the possibility that dynamics is more interesting than Newtonian dynamics
in a global asymptotically flat background. In the standard model, Newtonian
dynamics is na\"{\i}vely assumed to apply at the scale of galaxies and galaxy
clusters. However, this may not be the case. For example, Cooperstock and Tieu
\cite{CT1,CT2} have shown that stationary axisymmetric rotating dust
solutions\footnote{The solutions are circular: i.e.,
with zero expansion and shear, but nonvanishing vorticity.} may
be solved to phenomenologically match the rotation curves of certain
spiral galaxies, whose observed density distribution might be plausibly
approximated by circular symmetry (neglecting the density contrasts of
spiral arms or bars). Although various details of the Cooperstock--Tieu
model are debated \cite{CT2,RS}, it does demonstrate that the nonlinearity
of the Einstein equations is a potentially significant complicating feature
for extended matter distributions, even in the weak field limit.

\subsubsection{Galaxy cluster dynamics}
In proceeding to level (iv) -- galaxy clusters -- the fundamental issues
become obviously nontrivial. Since many galaxy clusters are spherical in shape
there is a temptation to model them using the spherically symmetric
dust Lema\^{\i}tre--Tolman--Bondi (LTB) solutions \cite{L,T,B}. While LTB
models have certainly been applied to structure formation \cite{KH1,KH2}, their
applicability is constrained by the uniform spherical shell approximation
remaining valid, without shell crossings or the growth of angular momentum
perturbations. This is probably unrealistically constraining for the case of a
generic collapse, and LTB models are most obviously applicable to expanding
spherical voids \cite {BKH} with ionic or molecular sized dust. If we consider
{\em virialized} spherical clusters of galaxies then there is no obvious
reason for the LTB model to be applicable. In many
galaxy clusters the motion of individual galaxies may be close to radial --
however, the phases of the galaxies relative to passage through the centre of
the cluster are completely uncorrelated. Individual galaxies will pass close to
the core of the cluster and emerge from the other side; at any instant the
number of galaxies moving out from the centre might be comparable to the number
falling in. Thus virialized galaxy clusters certainly do {\em not} have the
symmetry of a spherically symmetric dust solution if the galaxies are to be
identified as the dust.

The real question is: can we nonetheless model such systems as spherically
symmetric solutions of Einstein's equations with an effective purely radial
pressure, and possibly also effective heat flow terms? Or do we have to go
beyond spherically symmetric solutions to consider possible effective
anisotropic stresses? Since the interaction between individual galaxies in
a virialized cluster is purely gravitational, we see that this question
is really one of the statistical nature of gravity under
coarse--graining. Namely, for virialized systems with a manifestly spherical
shape, can the statistical properties of the individual gravitational
interactions of the dust particles (galaxies) be described by Einstein's
equation with a spherically symmetric effective fluid, or otherwise?

\subsubsection{Cosmological dynamics}
The final levels (v) and (vi) of coarse-graining in going to cosmological
averages involve qualitatively new fundamental questions. If we require that a
single model should describe the evolution of the universe from last scattering
to the present day, then we must coarse grain on scales over which the notion
of a dust ``particle'' has a meaning from last scattering up to the present
day. The description of a galaxy composed of stellar particles, or of a
virialized galaxy cluster composed of galaxy particles is only valid for those
epochs after which the relevant ``particles'' have formed and are themselves
relatively unchanging. Over cosmological timescales we do not have
well-defined invariant dust particles. The nature of galaxies and galaxy
clusters changes through growth by accretion of gas and by mergers.

The problem of cosmological dynamics is therefore essentially different to that
of the dynamics of galaxies or virialized galaxy clusters, as we can no longer
make a stationary approximation. In order to circumvent the problem of
ill-defined particle-like building blocks, an alternative strategy is
that we coarse-grain the dust on scales large enough that the {\em average}
flow of mass from one particle to another is negligible up to the present
epoch. Although galaxy clusters vary greatly in their size and complexity,
there are no common virialized structures larger than clusters. Thus any level
of coarse-graining on scales larger than clusters necessarily means dealing
with dust ``particles'' that are themselves {\em expanding}, i.e.,
with objects more akin to fluid elements in hydrodynamics. This feature
gives the first fundamental qualitative difference for the
cosmological problem as compared to that of galaxies or galaxy clusters.

Although we can receive signals from anywhere within our particle horizon,
if we make the reasonable assumption that the amount of matter absorbed
from cosmic rays from distant galaxies is negligible, then the region which
has contributed matter particles to define the local geometry of our own
galaxy is very small. This bounding sphere, which Ellis and Stoeger \cite{ES09}
call the {\em matter horizon}, is estimated by them to be of order $2\,$Mpc for
the Milky Way using assumptions about the growth of perturbations from the
standard cosmology. This scale coincides roughly with the scale at which the
Hubble flow is believed to begin in the immediate neighbourhood of the local
group of galaxies. It is one way of realising the concept of {\em finite
infinity}, introduced qualitatively by Ellis in his first discussion of the
fitting problem \cite{fit1}.

The second qualitative difference is that we have to deal with expanding
fluid elements that have vastly different densities at the present epoch, and
which evolve more or less independently. For galaxy clusters some sort of
finite infinity notion, with a variable scale of order $2$--$10\,$Mpc
depending on the size of cluster might be useful for defining independent
fluid elements. By combining such regions we arrive at the walls and filaments
that contain most of the mass of the universe. However, to this we must also
add the voids which dominate the volume of the universe at the present
epoch. These are the regions in which structures have never formed, and
which still contain the same ionic, atomic and molecular dust content
that has existed since very early epochs, only greatly diluted by expansion.

Any relevant averaging scale is therefore phenomenologically related to the
observed statistical distribution of void sizes. A precise definition of a void
fraction of course depends on the definition of a void. Surveys indicate that
voids with characteristic mean effective radii\footnote{Voids display a degree
of ellipticity. The {\em mean effective radius} of a void is that of a sphere
with the same volume as occupied by the void \cite{HV1,HV2,Pan11}, which is
typically larger than the maximal sphere enclosed by the same void.} of order
$(15\pm3)h^{-1}\,$Mpc (or diameters of order $30h^{-1}\,$Mpc), and a typical
density contrast of $\de\rh/\rh=-0.94\pm0.02$, make up 40\% of the volume of
the nearby universe \cite{HV1,HV2}. A very recent study \cite{Pan11} of the
Sloan Digital Sky Survey Data Release 7 finds a median effective void radius of
$17h^{-1}\,$Mpc, with voids of effective radii in the range $10h^{-1}\,$Mpc to
$30h^{-1}\,$Mpc occupying 62\% of the survey volume. In addition to these there
are numerous smaller minivoids \cite{minivoids}, which combined with the
dominant voids ensure that overall voids dominate the present epoch universe by
volume.

For the purposes of coarse-graining, what is important is not the overall
volume fraction of voids, but the size of the typical largest structures.
Any minimal scale for the cosmological coarse-graining of the final smoothed
density distribution has to be substantially larger than the diameter of the
largest structures. Void statistics \cite{Pan11} indicate an effective cutoff
of $60h^{-1}\,$Mpc for the largest mean effective diameters of voids, i.e.,
twice the scale of the typical dominant void diameters. Thus observationally,
the relevant scale for coarse-graining appears to be at least three times the
dominant void diameter, i.e., of order $100h^{-1}\,$Mpc, which coincides
roughly with the Baryon Acoustic Oscillation (BAO) scale.

\subsection{Approaches to coarse-graining}
The physical degrees of freedom which we must coarse grain are contained
in the curvature tensor and the sources of the field equations (\ref{Efe}).
In principle coarse-graining the curvature tensor might involve steps other
than simply coarse-graining of the metric. However, if a metric description of
gravity is assumed at each level, then schematically the hierarchy of
coarse-graining might be heuristically described as
\beq
\left. \begin{array}{r}
\metric{stellar}\to\metric{galaxy}\to\metric{cluster}\to
\metric{wall}\\ \vdots\quad \\ \metric{void} \end{array}
\right\}\to \metric{universe}
\label{coarse}\eeq
where the ellipsis denotes the fact that the metric of more than one type of
wall or void might possibly be relevant. In this scheme the lowest members
are assumed to be well modeled by exact solutions of Einstein's field
equations: $\metric{stellar}$ being a solution to the vacuum field equations
in a stellar system with a star or black hole as source (typically the
Schwarzschild solution or Kerr solution), and $\metric{void}$ being that of
a region filled with low density ionic dust with whatever symmetries are
relevant.

Dealing with this problem is obviously very complicated, and the
models that have been studied to date typically treat just one level in the
hierarchy.

\subsubsection{Discretized universes}\qquad
One of the few approaches to tackling the dust approximation head-on is
the Lindquist--Wheeler model \cite{LW}, which has received new interest
recently \cite{CF,Cl10,UEL}. In this approach the coarse--graining
hierarchy (\ref{coarse}) is replaced by the simplified scheme
\beq
\metric{stellar}\to \metric{universe}
\eeq
with the understanding that here $\metric{stellar}$ denotes the Schwarzschild
solution and could be taken as a substitute for either galaxies or clusters
of galaxies, depending on the masses assigned to the objects in the lattice.
Furthermore, here $\metric{universe}$ does not play a tangible geometrical
role. No cosmological metric is assumed in Einstein equations
at the outset. Rather, by matching the spherical boundaries of radially
expanding geodesics in the Schwarzschild geometries of a regular lattice of
equal point masses, the Friedmann equations are obtained \cite{LW,CF}. The
matching is exact only at the points where the radial spheres intersect and
is approximate in the regions in which spheres overlap or are excluded.

This model is analogous to the Swiss cheese models \cite{ES,MN11} in the
sense that the point group symmetry of the lattice is a discretized version
of overall global spatial homogeneity. The principal difference from
Swiss cheese models is that one is not cutting and pasting spheres into
a pre-existing continuum spacetime. Rather the continuum geometry is
only realised as an approximate description of the underlying discretized
space. Since the approximate continuum geometry is a FLRW model, the
Lindquist--Wheeler model has much in common with the models listed in section
\ref{FLRWa} below.  The symmetry of the lattice is such an integral part of
the construction that it is difficult to envisage how such models could be
easily generalized to more typically inhomogeneous cases. Nonetheless, it is
an extremely interesting toy model in which questions such as light
propagation can be investigated with detailed control.

Clifton and Ferreira have carefully studied the propagation of light in
a Lindquist--Wheeler model which approximates a spatially flat Einstein--de
Sitter model \cite{CF}. They find a significant deviation of the redshift,
$z$, of the lattice universe as compared to that of the FLRW universe,
$z\Ns{FLRW}$. For typical null geodesics, numerical calculations show
that $1+z\simeq(1+z\Ns{FLRW})^{7/10}$.
Essentially, this might be considered as a difference from the focusing
arising from Weyl curvature in the case of the point masses, as compared
with Ricci curvature focusing for a continuous dust fluid. While the
change in the luminosity distance--redshift relation is in the opposite
direction as compared to what is required for a viable explanation of the
expansion history of the universe without dark energy, the large difference
between the discrete and continuum cases demonstrates that we cannot
confidently claim to have reached an era of ``precision cosmology'' as long
as such fundamental issues as that of coarse-graining and the dust
approximation are not understood.

\subsubsection{Korzy\'nski's covariant coarse-graining\label{Kcoarse}}\qquad
Korzy\'nski \cite{Ko09} has recently proposed a covariant coarse-graining
procedure to be applied to dust solutions. This procedure could conceivably
be applied to any step in the hierarchy (\ref{coarse}) for which the starting
point is the metric of a known dust solution. Korzy\'nski also discusses the
special limit of replacing a dust world tube by a single worldline \cite{Ko09},
which might be viewed as proceeding in the opposite direction to that taken
in the Lindquist--Wheeler model.

Korzy\'nski's idea is to isometrically embed the boundary of a comoving
dust-filled domain -- required to have $S^2$ topology with positive scalar
curvature -- into a three-dimensional Euclidean space, and to construct
a ``fictitious'' three-dimensional fluid velocity which induces the same
infinitesimal metric deformation on the embedded surface as the ``true''
dust flow does on the domain boundary in the original spacetime. This
velocity field is used to uniquely assign coarse-grained expressions for the
volume expansion and shear to the original domain. An additional
construction using the pushforward of the ADM shift vector is used to
similarly obtain a coarse-grained vorticity. The coarse-grained quantities
are quasilocal functionals which depend only on the geometry of the
boundary of the relevant domain.

Korzy\'nski's approach represents an interesting new way of attacking
the fitting problem, and may also provide a useful framework for formulating
the problem of backreaction.

\subsection{Averaging and backreaction}

Averaging and coarse-graining are of course intimately related. The basic
distinction is that with averaging one is interested in the overall average
dynamics and evolution, most often without direct consideration of the details
of the course-graining procedure. Whereas coarse-graining is little studied,
much more attention has been paid to averaging, and a number of different
approaches have been pursued. These approaches are also discussed in the
reviews of van den Hoogen \cite{vdH}, Ellis \cite{E11} and Clarkson \etal\
\cite{CELU}.

Whereas coarse-graining is generally a bottom-up procedure, averaging is
top-down\footnote{The terms ``averaging'' and ``coarse-graining'' are often
used interchangeably in a loose sense. One might view averaging as a ``top-down
coarse-graining procedure''; just there are many ways of coarse-graining, and
here I have reserved the term for the bottom-up approaches.}
as it usually starts from the assumption that a well-defined average exists,
with a number of assumed properties. Generically, if one assumes that
the Einstein field equations (\ref{Efe}) are valid for some general
inhomogeneous geometry, $\metric{}$, then given some as yet unspecified
averaging procedure denoted by angle brackets, the average of (\ref{Efe})
gives
\beq
\ave{{G^\mu}_\nu}=\ave{g^{\mu\la}R_{\la\nu}}-\frn12{\de^\mu}_\nu\ave{g^{\la\rh}
R_{\la\rh}}={8\pi G\over c^4}\ave{{T^\mu}_\nu}\,.
\label{aEfe}
\eeq
At this point a number of choices are possible since there is no reason
to necessarily assume that  $\ave{{G^\mu}_\nu}$ is the Einstein tensor
of an exact geometry. In other words, on cosmological scales there is
no a priori necessity for (\ref{aEfe}) to correspond to an exact solution
of Einstein's equations.

In his ``macroscopic gravity'' approach, Zalaletdinov \cite{Z1,Z2,Z3} chooses
to work with the average inverse metric $\ave{g^{\mu\nu}}$ and the average
Ricci tensor $\ave{R_{\mu\nu}}$ and to write
\beq
\ave{g^{\mu\la}}\ave{R_{\la\nu}}-\frn12{\de^\mu}_\nu\ave{g^{\la\rh}}\ave{
R_{\la\rh}}+{C^\mu}_\nu={8\pi G\over c^4}\ave{{T^\mu}_\nu}\,,
\label{Zfe}
\eeq
where the correlation functions ${C^\mu}_\nu$ are defined by the difference
of the left hand sides of (\ref{Zfe}) and (\ref{aEfe}). Zalaletdinov
provides additional mathematical structure to prescribe a covariant averaging
scheme, thereby defining properties of the correlation functions.

Another way of formulating the problem is to work in terms of the difference
between the general inhomogeneous metric and the averaged metric
\beq
\metric{}=\bar\metric{}+\de\metric{},\qquad
\label{metdef}
\eeq
where $\bar\metric{}\equiv\ave{\metric{}}$, with inverse
$\bar g^{\la\mu}\ne\ave{g^{\la\mu}}$. We can now determine a connection
$\bar\GA^\la_{\ \mu\nu}$, curvature tensor ${\bar R^\mu}_{\ \;\nu\la\rh}$
and Einstein tensor $\bar G^\mu_{\ \nu}$ based on the averaged metric,
$\bar\metric{}$, alone.
The differences $\de\GA^\la_{\ \mu\nu}\equiv\ave{\GA^\la_{\ \mu\nu}}-\bar\GA^
\la_{\ \mu\nu}$, $\de R^\mu_{\ \nu\la\rh}\equiv\ave{R^\mu_{\ \nu\la\rh}}-\bar
R^\mu_{\ \nu\la\rh}$, $\de R_{\mu\nu}\equiv\ave{R_{\mu\nu}}-\bar R_{\mu\nu}$
etc, then represent the {\em backreaction} of the average inhomogeneities on
the average geometry determined from $\bar\metric{}$. Furthermore, the average
Einstein field equations (\ref{aEfe}) may be written
\beq
\bar G^\mu_{\ \nu}+\de G^\mu_{\ \nu}={8\pi G\over c^4}\ave{{T^\mu}_\nu}\,.
\label{bEfe}
\eeq
This expresses the fact that the Einstein tensor of the average metric
is not in general the average of the Einstein tensor of the original metric.
The processes of averaging and constructing the Einstein tensor do not
commute.

Equations (\ref{aEfe}) and (\ref{bEfe}) are of course very similar, but
may differ in both the overall average represented by the angle brackets,
and also in the split between the background averaged Einstein tensor and
the correlation or backreaction terms.

There are three main types of averaging schemes that have been considered.
They can be classed as:
\begin{itemize}
\item Perturbative schemes about a given background geometry;
\item Spacetime averages;
\item Spatial averages on hypersurfaces based on a $1+3$ foliation of
spacetime.
\end{itemize}
I will briefly describe each case in turn.

\subsubsection{Perturbations about exact cosmological spacetimes}

A vast literature exists on inhomogeneous models treated as perturbations
of the exact FLRW models. In this approach one assumes that the average
geometry $\bar\metric{}$ of (\ref{metdef}) is exactly a FLRW model, and
the quantities $\de\metric{}$ are to be treated as perturbative corrections.

The issue of whether backreaction is significant or insignificant in the
perturbative FLRW context is a matter of much debate, with different authors
coming to different conclusions, which may be traced to various differences
in assumptions made. Since these issues are discussed in many other reviews,
such as those of Clarkson \etal\ \cite{CELU}, Kolb \cite{K11} and the paper of
Clarkson and Umeh \cite{C11}, I will not discuss the perturbative approach in
detail here.

The perturbative approach of course relies on the assumption that a FLRW
model exactly describes the average evolution of the universe at the largest
scales, and this may be incorrect. Related physical issues are further
discussed in section~\ref{Machh}.

\subsubsection{Spacetime averages: Zalaletdinov's macroscopic gravity
\label{secZ}}

General covariance is generally seen as a desirable property, since it
is an essential feature of general relativity that physical quantities
should not depend on arbitrary choices of coordinates. However, any process
of taking an average will in general break general covariance. Furthermore,
if an average geometry on cosmological scales no longer satisfies the Einstein
equations, which is a distinct possibility given that solutions of (\ref{Efe})
are only directly tested on the scale of stellar systems, then the role that
general covariance plays in defining spacetime structure on the largest
scales may need to be revisited from first principles. In particular,
although we might still desire that physical quantities should not depend
on choices of coordinates, the relationship of the coordinates of a
``fine--grained manifold'' relative to those of an average ``coarse--grained
manifold'' need to be carefully considered.

Zalaletdinov views general covariance as paramount in determining macroscopic
spacetime structure, and he introduces additional mathematical structure to
perform averaging of tensors in a covariant manner on a given manifold, $\Mm$
\cite{Z1,Z2,Z3}. His aim is to consistently average the Cartan equations from
first principles, in analogy to the averaging of the microscopic
Maxwell--Lorentz equations in electromagnetism. However, whereas
electrodynamics is linear in the fields on the fixed background of Minkowski
spacetime, gravity demands an averaging of the nonlinear geometry of spacetime
itself.

The additional structure introduced by Zalaletdinov \cite{Z1,Z2,Z3} takes the
form of bilocal averaging operators, ${\Az^\mu}_{\al}(x,x')$, with support at
two points $x\in\Mm$ and $x'\in\Mm$, which allow one to construct a bitensor
extension, ${\Tz^\mu}_\nu(x,x')$, of a tensor ${T^\mu}_\nu(x)$ according to
\beq
{\Tz^\mu}_\nu(x,x')={\Az^\mu}_{{\al'}}(x,x'){T^{\al'}}_{{\be'}}(x')
{\Az^{\be'}}_{\nu}(x',x)\,.
\eeq
The bitensor extension is then integrated over a 4-dimensional spacetime
region, $\SI\subset\Mm$, to obtain a regional average according to
\beq
\bar T^\mu_{\ \nu}(x)={1\over\Vv_\SI}\int_\SI\dd^4 x'\,\sqrt{-g(x')}\,
{\Tz^\mu}_\nu(x,x'),
\eeq
where $\Vv_\SI\equiv\int_\SI\dd^4 x\,\sqrt{-g(x)}$ is the spacetime volume
of the region $\SI$. The bitensor transforms as a tensor at every point but is
a scalar when integrated over a region for the purpose of averaging.

In the macroscopic gravity approach, much effort has been expended
\cite{Z1,Z2,Z3,MZ} in developing a mathematical formalism which in the average
bears a close resemblance to general relativity itself. Indeed, apart from the
fact that the macroscopic scale is assumed to be larger than the microscopic
scale, there is no scale in the final theory. As Clarkson \etal\ \cite{CELU}
have already commented, this is also potentially a weakness of the macroscopic
gravity approach. Observations suggest a complex hierarchy of averaging, given
by the scheme (\ref{coarse}), which may involve physical issues more complex
than simply one step from a microscopic theory to a macroscopic theory of
gravity. Indeed, a number of the steps associated with the observed scales of
coarse-graining phenomenologically involve going from background solutions of
Einstein's field equations {\em with particular symmetries} to other solutions
of Einstein's field equations {\em with particular symmetries}. Therefore it is
a highly nontrivial question as to whether the physically relevant mathematical
framework is one which takes us from one version of a diffeomorphism invariant
theory of gravity with no specific symmetries to another diffeomorphism
invariant theory of gravity with no specific symmetries, which is precisely
what Zalaletdinov has constructed. Specific cosmological questions involve the
choice of specific macroscopic scales.

In practice, cosmological applications of Zalaletdinov's formalism have
involved making additional assumptions, such as those which lead to a
spatial averaging limit \cite{PS07}, or additionally in {\em assuming} that
the average geometry is a FLRW geometry \cite{CPZ,P08,PS08,vdH09}. In this
case it is found that the macroscopic gravity correlation terms
take the form of a spatial curvature, even though a spatially {\em flat} FLRW
geometry is assumed for the average geometry \cite{CPZ}.

Recently an alternative covariant spacetime averaging scheme has been put
forward by Brannlund, van den Hoogen and Coley \cite{BvdHC}. It treats the
manifold as a frame bundle as a starting point for the averaging of geometric
objects.

\subsubsection{Spatial averages: Buchert's formalism\label{secB}}

Building on earlier work \cite{BE,EB,CP} in the late 1990s Buchert
developed an approach \cite{buch1,buch2} for the spatial averaging of scalar
quantities associated with the Einstein field equations (\ref{Efe}) in the
$1+3$ ADM formalism, with cosmological averages in a fully nonperturbative
relativistic setting in mind at the outset. The $1+3$ setting is natural if the
Einstein field equations (\ref{Efe}) are to  be viewed as evolution equations.

Rather than introducing additional structure to fully tackle
the mathematically difficult problem of averaging tensors, Buchert
approached the problem by just averaging scalar quantities associated with
spacetimes with inhomogeneous perfect fluid energy--momentum sources.
Such scalars include the density, $\rh$, expansion, $\th$, and scalar shear,
$\si^2=\frn12\si_{\al\be} \si^{\al\be}$ etc. For an arbitrary manifold, ADM
coordinates
\beq\ds^2=-\Bom^0\otimes\Bom^0+g_{ij}(t,\bx)\,\Bom^i\otimes\Bom^j,
\label{split}\eeq
where
\bea\Bom^0&=&\NN(t,\bx)\,c\,\dd t,\nonumber\\
\Bom^i&=&\dd x^i+\NN^i(t,\bx)\,c\,\dd t.\label{shift}
\eea
can always be chosen locally but not necessarily globally. Buchert restricted
the evolution problem to that of irrotational flow in order that (\ref{split}),
(\ref{shift}) can be assumed to apply over global $t=\,$const spatial
hypersurfaces. For a dust source\footnote{The extension to perfect fluids was
introduced in \cite{buch2}, and to other matter sources in \cite{BBR}. Further
extensions that deal with general hypersurfaces tilted with
respect to the fluid flow have been discussed by various authors
\cite{L09}--\cite{R10}.} we can then choose synchronous coordinates with
$\NN=1$ and $\NN^i=0$. With these choices, the Einstein equations may be
averaged on a domain, $\DD$, of the spatial hypersurfaces, $\SI$, to give
\bea
3{\dot\ab^2\over\ab^2}&=&8\pi G\ave\rh-\half c^2\Rav-\half\QQ,\label{buche1}\\
3{\ddot\ab\over\ab}&=&-4\pi G\ave\rh+\QQ,\label{buche2}\\
\pt_t\ave\rh&+&3{\dot\ab\over\ab}\ave\rh=0,
\label{buche3}\eea
where the overdot denotes a $t$--derivative,
\beq
\QQ\equiv\frac23\left\langle\left(\th-\langle\th\rangle\right)^2\right\rangle
-2\ave{\si^2}
=\frac23\left(\langle\th^2\rangle-\langle\th\rangle^2\right)-
2\ave{\si^2}\,,
\label{backr}\eeq
is the {\em kinematic backreaction}, and angle brackets denote the spatial
volume average of a quantity, so that $\Rav\equiv\left(\int_\DD\dd^3x\sqrt
{\det\gd}\,{\cal R}(t,\bx)\right)/\Vav(t)$ is the average spatial curvature,
for example, with $\Vav(t)\equiv\int_\DD\dd^3x\sqrt{\det\gd}$ being the volume
of the domain $\DD\subset\SI$. It is important to note that $\ab$ is {\em not}
the scale factor of any given geometry, but rather is defined in terms of the
average volume according to
\beq
\ab(t)\equiv\left[\Vav(t)/\Vav(t\Z0)\right]^{1/3}\,.
\eeq
It follows that the Hubble parameter appearing in
(\ref{buche1})--(\ref{buche3}) is, up to a factor, the  volume average
expansion scalar, $\th$,
\beq
{\dot\ab\over\ab}=\frn13\ave{\th}.\label{Hav}
\eeq
The following condition
\beq \pt_t\left(\ab^6\QQ\right)+\ab^4c^2\pt_t\left(\ab^2\Rav\right)=0,
\label{intQ}\eeq
is required to ensure that (\ref{buche1}) is the integral of (\ref{buche2}).

In Buchert's scheme the non-commutativity of averaging and time evolution is
described by the exact relation \cite{BE,EB,buch1,RSKB}
\beq\pt_t{}\ave\Psi-\ave{\pt_t\Psi}=\ave{\Psi\th}
-\ave\th\ave\Psi\label{comm}\eeq
for any scalar, $\Psi$.

The operational interpretation of Buchert's formalism poses many questions,
which we will return to in section~\ref{obss}. Leaving these issues
aside, (\ref{buche2}) is already suggestive since it implies that if
the backreaction term is large enough -- for example, in the case of a
large variance in expansion rate with small shear -- then the volume average
acceleration, $\ab^{-1}\ddot\ab=\frn13\Der\dd t\ave\th+\frn19\ave\th^2$,
could be positive, even if the expansion of all regions is locally
decelerating. Essentially, the fraction of the volume occupied by the
faster expanding regions is initially tiny but may become significant at
late epochs, skewing the average to give an illusion of acceleration
during the period in which the voids start to dominate the volume average.
Whether this is observationally viable, however, depends crucially on how
large the variance in expansion rates can grow given the initial constraints
on density perturbations, and on the operational interpretation of the
Buchert average.

Another big question is the extent to which the truncation of the averaging
problem from the full Einstein equations to scalar evolution equations can be
derived from a more fundamental basis. If density perturbations are the most
important phenomenologically, then the Buchert scheme may well be justified,
but it then needs to be understood as an appropriate limit in a more general
scheme. Coley \cite{Co10} suggests that since 4-dimensional Lorentzian
manifolds can be completely characterized by their scalar polynomial curvature
invariants, this might provide a suitable mathematical basis for an
averaging scheme based on scalars. In practice, however, we are still
faced with the same observational interpretation problems when dealing
with a single null cone average versus spatial averages of statistical
ensembles.

Korzy\'nski's covariant coarse-graining approach \cite{Ko09}, discussed in
section~\ref{Kcoarse}, when applied to irrotational dust might be viewed as a
generalization of the Buchert approach, which gives rise to additional
backreaction terms. The extent to which Buchert's scheme might be viewed as
a limit of Korzy\'nski's approach remains to be determined.

Within the context of $1+3$ formalisms, there have been a number of
studies of averaging and backreaction which focus principally on associated
mathematical issues. These include Ricci flow \cite{CP,CM,BC1,CB}, group
averaging of the FLRW isometry group \cite{A09} and the characterization of
constant mean (extrinsic) curvature (CMC) flows \cite {Re07,Re08,Re09}. Such
approaches might provide further insights into the general problem of
averaging tensors. For example, the Ricci flow is a well studied procedure in
Riemannian geometry which may be used to realize a regional smoothing through
a rescaling of the metrical structure \cite{CP,CM,BC1,CB}, in the spirit of
renormalization group flows. Since the primary motivation of this review is
physical, I will not further discuss these approaches here.

\section{Average spatial homogeneity: How do we define it?}

The very near isotropy of the CMB demonstrates that when photons travel
to us from the surface of last scattering, then to a very good approximation
the geometry of the universe must be isotropic in some average sense. If
we assume some sort of statistical Copernican principle, then
we can also expect some sort of average notion of spatial homogeneity. The
hard question is: how do we convert the observed averaged isotropy of
the geometry on our past light cone into an appropriate notion of average
spatial homogeneity?

It is my own view that whereas a lot of effort has been expended in
defining the mathematics of averaging, not enough attention has been
given to the foundational physics underlying the appropriate notion of
an average. I will outline my views as to the best way to proceed in
section~\ref{Machh}, but will first describe the two approaches that have
received the most attention.

\subsection{The Friedmann--Lema\^{\i}tre universe as the average\label{FLRWa}}

The remarkable success of the standard cosmology, albeit with sources of
dark matter and dark energy for which there is no direct evidence on the
scale of the solar system, understandably leads most researchers to
assume that it must be correct, even if only in an average sense.

As a consequence, even when researchers study inhomogeneity then, putting
aside exact inhomogeneous solutions, the FLRW
model is simply {\em assumed} as the average in a majority of approaches. A
partial list of models for which this is the case includes:
\begin{itemize}
\item All perturbative calculations about the FLRW universe (whether
based on the standard \LCDM\ cosmology or otherwise);
\item any LTB models for which the universe is ``asymptotically FLRW''
with a core spherical inhomogeneity;
\item the Dyer-Roeder approach \cite{DR};
\item Swiss cheese \cite{ES} and meatball \cite{MN11,PGMK} models;
\item studies of spatial averaging in a $(1+3)$ setting which
derive more general mathematical results, but then assume the FLRW model
as the average when drawing specific conclusions in a cosmological context
\cite{RSKB,GW};
\item studies based on Zalaletdinov's macroscopic gravity which
derive more general mathematical results, but then assume the FLRW model
as the average when drawing specific conclusions in a cosmological context
\cite{PS07,CPZ,P08,PS08,vdH09};
\item studies of CMC flows which
derive more general mathematical results, but then assume the FLRW model
as the average when drawing specific conclusions in a cosmological context
\cite{Re07,Re08}.
\end{itemize}

I will not deal further with the details of these approaches, since several
other papers in this special issue deal with them. The main comment I wish to
make is that at the point that the FLRW model is introduced these approaches
effectively assume that on large enough scales the average geometry is
described by Einstein's equations (\ref{Efe}) with a spatially homogeneous
perfect fluid source, or more specifically a spatially homogeneous
dust source if we consider late epoch cosmic evolution. In other words,
although many of the general results derived in some of these approaches
may be quite broadly applicable, assumption of the FLRW average
demands very specific properties of dust in the unsolved processes
of coarse-graining discussed in section~\ref{grain}. Since the large scale
averages involve coarse--graining on scales on which space is expanding,
it generally means extrapolating the dust approximation to scales on
which usual notions of dust particles as bound systems cannot apply.

Furthermore, the notion of an average that these approaches implicitly demand
is also very restrictive, since it involves (at least) three conditions:
\begin{enumerate}
\item The notion of average spatial homogeneity is described by a class of
ideal comoving observers with synchronized clocks.
\item The notion of average spatial homogeneity is described by average
surfaces of constant spatial curvature (orthogonal to the geodesics
of the ideal comoving observers).
\item The expansion rate at which the ideal comoving observers separate within
the hypersurfaces of average spatial homogeneity is uniform.
\end{enumerate}

Already at the level of perturbation theory about FLRW models, one can
specialize to spacetime foliations which preserve one of the notions (i)--(iii)
of average spatial homogeneity more fundamentally than the other two, as was
already discussed many years ago in the classic work of Bardeen \cite{B80}.
Since spatial curvature is not specified by a single scalar in general, there
are many ways of realizing spacetime foliations which preserve one notion of
average spatial homogeneity more strongly than the others. Among the foliations
discussed by Bardeen we can recognise those of each type above: the
{\em comoving hypersurfaces} (and related synchronous gauge) take property (i)
as more fundamental; the {\em minimal shear hypersurfaces}\footnote{For scalar
perturbations this becomes a zero--shear condition, i.e., $\K_{ij}-\frn13g_{ij}
\K=0$, where $\K_{ij}$ is the extrinsic curvature, $g_{ij}$ the intrinsic
metric, and $\K\equiv{\K^\ell}_\ell$. For general perturbations the
hypersurfaces are defined by $\left(\K_{ij}-\frn13g_{ij}\K\right)_{|ij}=0$,
where the bar denotes a covariant derivative with respect to the intrinsic
metric.} (and related Newtonian gauge) are one type of foliation for which
property (ii) is more fundamental; and finally the {\em uniform Hubble flow
hypersurfaces} take property (iii) as more fundamental.

The possible foliations of perturbed FLRW models were recently considered in
considerable detail by Bi\v{c}\'ak, Katz and Lynden-Bell \cite{BKL07}, with a
view to making gauge choices that provide a realization of Mach's principle, in
the sense that the rotations and accelerations of local inertial frames can be
determined directly from local energy--momentum perturbations $\de{T^\mu}_\nu$.
The choices of hypersurfaces they consider are: uniform Hubble flow
hypersurfaces; uniform intrinsic scalar curvature hypersurfaces; and minimal
shear hypersurfaces. The {\em uniform intrinsic scalar curvature hypersurfaces}
provide a foliation in addition to those considered by Bardeen, which also take
property (ii) as more fundamental. In  addition to a choice of hypersurface
Bi\v{c}\'ak, Katz and Lynden-Bell further fix the gauge by an adopting a
condition similar to the ``minimal shift distortion condition'' of Smarr and
York \cite{SY}. With this condition, for each choice of hypersurface the
coordinates of local inertial frames are more or less uniquely determined by
the energy--momentum perturbations $\de{T^\mu}_\nu$. These ``Machian gauges''
are therefore substantially more restrictive than the commonly used synchronous
gauge and generalized Lorenz--de Donder gauge \cite{BKL07}.

\subsection{Constant time hypersurfaces as the average}

If we abandon the assumption that the average notion of spatial homogeneity is
given by an exact solution of Einstein's field equations with a perfect fluid
source, then there is no reason to assume that all of the conditions (i)--(iii)
described in the last section need to apply. In general, we need just one
condition to characterize the average; the question is which one?

Perhaps for the same historical reasons that led to the popularity of
the synchronous gauge, the most studied choice of spacetime split beyond
the perturbative regime is that of constant time hypersurfaces
orthogonal to ideal observers ``comoving with the dust'', even though the
nature of the dust is not generally prescribed. The Buchert approach to
spatial averaging \cite{buch1,buch2} grew as a generalization of averaging
in Newtonian cosmology \cite{BE,EB}. Since the split of space and time is
unique in Newtonian theory, from the Newtonian viewpoint this is the only
natural choice one can make.

If the particles of dust were observationally identifiable and invariant
from the time of last scattering until today, then there would be no physical
ambiguity about the notion of ``comoving with the dust''. In such a case,
the choice of constant time hypersurfaces with a synchronous gauge would be
well motivated. However, as discussed in section~\ref{grain}, in order to
consistently deal with both the particles of ionic dust in voids, and
also with ``particles'' of dust larger than galaxies, we have to coarse-grain
over fluid elements which are themselves expanding. This demands a detailed
understanding of the general statistical nature of general relativity in
the presence of complex sources, which is as yet unavailable. What is
true of Einstein's theory is that slicings based on any fixed
time or space cannot be expected to be the most natural choice of average
background. In my view, to understand the formidable interlocking problems
of averaging and the statistical nature of general relativity, we must
go back to first principles.

\subsection{Mach's principle, the equivalence principle, and the
average\label{Machh}}

In this section I will outline my views about an alternative physically
motivated approach to defining average spatial homogeneity
\cite{equiv1,equiv2}, which underlies the timescape cosmology
\cite{clocks,sol,obs}. Whether or not the current version of the timescape
cosmology is observationally viable as an alternative to the standard \LCDM\
cosmology without dark energy, the particular questions I wish to raise
are key ones which must be fully understood if we are to make progress with
the averaging problem.

In formulating general relativity as a dynamical theory of spacetime,
Einstein was guided philosophically by Mach's principle -- namely, the
broad notion that spacetime does not have a separate existence from
the material objects that inhabit it, but is a relational structure
between things. As Einstein put it \cite{esu}: {\em``In a consistent
theory of relativity there can be no inertia relatively to `space', but
only an inertia of masses relatively to one another''}. This after
all is the physical principle that underlies general covariance: there
is no absolute space or time, and so the basic laws of physics should not
depend on arbitrary choices of coordinates.

Einstein did not, however, fully succeed in implementing Mach's principle
in general relativity, since he never solved the global problem of
uniquely determining the structure of spacetime on large scales. As things
stand, his equations admit many cosmological solutions such as general
Bianchi models or the G\"odel universe, which do not look in the least
like the universe we actually inhabit. Indeed, such solutions might be
viewed as running counter to the spirit of Mach's principle. From the
cosmological viewpoint Mach's principle may be phrased \cite{Bondi}:
{\em``Local inertial frames are determined through the distributions of
energy and momentum in the universe by some weighted average of the apparent
motions''}. Although it is clear from this statement that any attempt
to tackle the averaging problem in cosmology must necessarily deal with
the issue of Mach's principle, relatively few authors
\cite{BKL07,equiv1,equiv2,BKL95,Sch1,Sch2} have approached the averaging
problem in these terms.

In considering the averaging problem, we should take
the underlying physical principles of relativity as a guide
to constructing the correct mathematical formalism, rather than simply trying
to mimic mathematical properties such as general covariance, which might
be relevant for formulating the microscopic nongravitational laws of physics
but are not necessarily relevant for very large scale averages of
gravitational degrees of freedom.

To prescribe the rules for defining spacetime in relational terms, it
is necessary to define the relationship between inertial frames and any
relevant mathematical structure. The strong
equivalence principle (SEP) stands as the concrete legacy of Einstein's
attempts to come to grips with Mach's principle, and it is the physical
cornerstone of general relativity. By the SEP the nongravitational laws
of physics should reduce to those of special relativity in local inertial
frames (LIFs) in the neighbourhood of an arbitrary spacetime point.
The principle of general covariance is a means of formulating the
nongravitational laws to achieve this.

However, when it comes to averages on large scales over which nongravitational
fields are negligible and only gravitational degrees of freedom prevail, we
are still left with the problem of defining a suitable ``weighted average of
the apparent motions''. To be consistent with the broad principles of
relativity, such an average must be limited by initial conditions and
causality. If Einstein's field equations (\ref{Efe}) are viewed as evolution
equations which determine the geometry dynamically, then causality should limit
the geometry at any event to only depend on the geometry within the past light
cone of all possible observers at that event. For realistic large scale
cosmological applications, given that energy absorbed from null signals
provides a negligible contribution as compared to local matter densities, it is
the local matter horizon \cite{ES09} which is actually the most relevant
domain in determining local average geometry.

In the perturbative FLRW framework, Bi\v{c}\'ak, Katz and Lynden-Bell
\cite{BKL07} have identified choices of hypersurfaces and coordinates within
those hypersurfaces, which are most uniquely restricted in terms of being
determined by local energy--momentum perturbations $\de{T^\mu}_\nu$ and thereby
represent ``Machian gauges''. If the average geometry is not an exact FLRW
model, then in going beyond the perturbative regime the question is which of
these notions best embodies Mach's principle?

\subsubsection{The cosmological equivalence principle}\qquad
I have argued \cite{equiv1,equiv2} that a nonperturbative generalization
of the ``uniform Hubble flow'' \cite{B80,BKL07} or CMC \cite{Y2,Y3}
slicing is the best choice for a further refinement of the notion of
inertia consistent with Mach's principle. The reader is referred to the essay
version \cite{equiv2} for a first introduction to these ideas, which I shall
only very briefly sketch here.

Since gravity sourced by matter obeying the strong energy condition is
universally attractive, any solution of Einstein's field equations (\ref{Efe})
on scales which only involve gravity is necessarily dynamical. In my view there
is a further principle of relativity that is lacking in general relativity with
the SEP alone, which is a consequence of taking this dynamical nature as
fundamental to the relational structure. In particular, for the regionally
homogeneous and isotropic volume--expanding part of geodesic deviation {\em it
should be impossible to distinguish whether particles are at rest in an
expanding space or alternatively are moving in a static Minkowski space}.

The relation to inertia can be understood in terms of the semi--tethered
lattice thought experiments \cite{equiv1,equiv2} in which a uniformly
expanding lattice of observers in Minkowski space, connected by freely spooling
isotropic tethers, apply brakes to the tethers with an impulse which is the
same function of the synchronized local proper time at each lattice site.
Deceleration takes place and work is done in converting the kinetic
energy of expansion to heat in the brakes. However, since the magnitude of
the impulse on each tether is identical at each lattice site, isotropy
guarantees that there is {\em no net force} on any observer of the lattice,
and so the motion is inertial in that sense, although a deceleration and
conversion of energy has taken place. In the gravitational case the regionally
homogeneous isotropic part of the density plays the same role as the brakes on
the tethers.

From the point of view of the averaging problem, my proposal is to restrict
the global geometry in the final step of the average (\ref{coarse}) to
one which can be decomposed into average domains which
obey the {\em cosmological equivalence principle} (CEP) \cite{equiv1}:
{\em At any event, always and everywhere, it is possible to choose a suitably
defined average spacetime region, the cosmological inertial frame (CIF), in
which average motions (timelike and null) can be described by geodesics in a
geometry that is Minkowski up to some time-dependent conformal
transformation},
\beq \ds^2\Ns{CIF}=
a^2(\et)\left[-\dd\et^2+\dd r^2+r^2(\dd\th^2+\sin^2\th\,\dd\ph^2)\right].
\label{cif}\eeq

While this statement of the CEP would reduce to the standard SEP if $a(\et)$
were constant, or alternatively over very short time intervals during which
the time variation of $a(\et)$ can be neglected, it is important to realize
that the averaging region represented by the CIF (\ref{cif}) is very much
larger than the neighbourhood of a point as assumed in a LIF. Relative to
bound systems, the spatially flat FLRW metric (\ref{cif}) is to be viewed
as applicable only on scales larger than galaxy clusters which correspond
to finite infinity \cite{fit1,clocks} or the matter horizon \cite{ES09}.
Alternatively, within void regions, which might be regionally modeled by a
portion of an open FLRW universe, a CIF would be applicable only on spatial
scales which are small with respect to the scalar curvature radius.

Although the spatial extent of a CIF would be much smaller within void regions
than within walls, it is important to recognise that unlike a LIF (\ref{cif})
is intended to apply on arbitrarily long timescales which capture the
volume--decelerating part of the average geodesic deviation. Rather than being
a geometry in the neighbourhood of a point, is an average asymptotic geometry
for spatial regions of order 1--10$\,$Mpc. On these scales (\ref{cif}) provides
a suitable geometry to replace the usual notion of an asymptotically flat
geometry for isolated systems. Although it is a spatially flat FLRW metric, it
is not a global geometry for the whole universe as is the case in the standard
cosmology.\setcounter{footnote}{1}

A more detailed discussion of the rationale behind the CEP, including the
roles of Weyl curvature and Ricci curvature in the averaging process, is
provided in \cite{equiv1}. The key idea is that
the CIF isolates a notion of inertia that only exists as a result of a {\em
collective} degree of freedom of the regional background. In a sense we are
dealing with the conformal mode of the 3--geometry which has been identified
before in discussions that attempt to isolate the true gravitational degrees of
freedom \cite{Y2,Y3,Y1} and related discussions of Mach's
principle\footnote{The separation of the conformal degree of freedom of the
3--metric from shape degrees of freedom is at the heart of Barbour's Machian
approach to gravity \cite{BB,ABFKO}, which he has approached from many angles
including the N-body problem \cite{B02}.} \cite{ABFKO}.

It is well--known that for the exchange of photons between comoving observers
in the background (\ref{cif}), to leading order the observed redshift
of one comoving observer relative to another yields the same local Hubble law,
whether the exact relation, $z+1=a\Z0/a$, is used or alternatively the radial
Doppler formula, $z+1=[(c+v)/(c-v)]^{1/2}$, of special relativity is used,
before making a local approximation. Making this a feature of regional
averages allows for forms of inhomogeneity that admit such an
indistinguishability of whether ``particles are moving'' or ``space is
expanding'', while disallowing global coherent anisotropic flows of the
sort typified by Bianchi models. Bianchi models single out preferred directions
in the global background universe, thereby imbuing spacetime with absolute
qualities that go beyond an essentially relational structure. To make general
relativity truly Machian such backgrounds need to be outlawed by principle,
and the CEP is one means to achieving this while allowing inhomogeneity.

To combine such regional average CIFs requires something akin to the
introduction of a CMC slicing to preserve a uniform Hubble flow condition.
By the SEP the first derivatives of the metric can always be set to zero
in the neighbourhood of a point; it is not the connection that corresponds
to the physical observables but the curvature tensor which is derived from
it. The possibility of always being able to choose a uniform Hubble flow
slicing extends this to regional scales -- the first derivatives of the
regional metric which correspond to the volume expansion are a degree of
freedom upon which physical observables do not depend. The Hubble parameter
is thus recognized as a ``gauge choice'' that can be made within the limits
set by evolution from initial conditions at early epochs. We are always
allowed to make a choice of coordinates of the averaging regions which keeps
the Hubble parameter uniform in expanding regions despite the presence of large
variations in regional densities and curvature. 

The mathematical procedures required to construct such a uniform flow slicing
in terms of patching one CIF region to the next have not been developed yet.
Since one is dealing with a nonperturbative regime without prescribed dust the
problem is likely to require mathematical constructions which go beyond the
treatment of CMC foliations in a globally well-defined background with a
prescribed energy-momentum tensor. The framework should involve a statistical
description of geometry closely related to the coarse-graining procedure.

\section{Cosmic averages versus cosmic variance\label{obss}}

In any description of cosmic averages we must ask the question of how
local observables are to be related to average quantities. There are
two aspects to this question:
\begin{enumerate}
\item How do our own measurements relate to some canonically defined
average quantity?
\item How do statistically average quantities defined on spatial slices
that define average cosmic evolution relate to average distances and angles
on the past light cone from which all cosmological measurements are inferred?
\end{enumerate}

These are significant questions which must be answered to build a viable
cosmological model. It was recognized early on in the Buchert approach that
this is a nontrivial problem, so that in general observed cosmological
parameters will be ``dressed'' when compared to the bare cosmological
parameters of the averaging scheme \cite{BC1,BC2}. However, additional
assumptions are required to achieve such a dressing. Rather than tackling this
problem, a common approach has been simply to identify the volume average
expansion parameter (\ref{Hav}) in the Buchert scheme with our observed Hubble
parameter \cite{morph1}--\cite{WB}, and to identify the observed redshift
according to $1+z=\ab\Z0/\ab$. However, such an assumption remains to be
rigorously justified.

In my view it is important that we think carefully about these questions
since once there is inhomogeneity and a variance in geometry, then not
every observer can be the same average observer. Understanding and quantifying
our observed measurements in relation to cosmic variance is as fundamentally
important as understanding cosmic averages.

Structure formation provides us with a natural separation of scales which
enable us to attack this problem from first principles \cite{clocks}. In
particular, we and all the galaxies we observe are bound structures which
necessarily formed from density perturbations that were greater than critical
density. Yet the volume of the universe is dominated at the present epoch by
voids. Thus an average position by volume -- which is operationally what the
Buchert average defines -- will be located in a void unbound to any structure,
in a region whose local density and spatial curvature differs markedly from
those in a galaxy where the actual objects we observe are located. The mass
average therefore does not coincide with the volume average, and there can be
systematic differences between the geometry of galaxies and the average
geometry, which must be taken into account.

We are therefore led to a {\em statistical Copernican principle}: we are
observers in an average galaxy, and in this sense our position is not special.
However, by virtue of being in a galaxy our local environment is not the
same density as the local environment of an average position by volume and
this fact has to be taken into account in our interpretation of cosmological
measurements. By analogy, in the original heliocentric solution Copernicus
realised that we are not in the centre of the universe, but the fact that
we are on the surface of a planet that rotates means our view of the universe
is different from that at a random point in the solar system, and must be
taken into account when interpreting astronomical observations.

The averaging problem has been studied in the Buchert formalism in various
approaches which partition the universe into regions of different density
\cite{clocks,sol,obs,R06,BC3,WB}. I will discuss the timescape cosmology in
particular, since among the various approaches it is unique in focusing on
the importance of the position of the observer in relation to average cosmic
evolution, rather than the average cosmic evolution alone. It provides a
prescription for quantifying the apparent variance in the expansion rate, and
for the interpretation of volume average cosmological quantities by observers
in galaxies, whose local geometry does not coincide with the volume-average
metric on account of the mass-biased selection effect.

\subsection{The timescape model and the Buchert formalism}
To date the timescape model \cite{clocks,sol,obs} has been developed by
adapting the Buchert formalism. Since the timescape model is based on the
idea that average spatial homogeneity is related to a uniform Hubble
flow condition, the use of the Buchert formalism -- with its choice of
comoving hypersurfaces and a synchronous gauge -- may seem contradictory.
However, dust is not prescribed in the Buchert scheme, and the timescape
model does not apply Buchert averaging to exact solutions with prescribed
dust.

The timescape model begins from the premise that dust is coarse-grained at
the $100h^{-1}\,$Mpc scale\footnote{As we will discuss in section~\ref{obsd},
the specification of length scales depends on the average metric description
any observer chooses. Here I assume a normalization of spatial distances to
a spatially flat metric, in accord with the conventions of the standard
cosmology.} of statistical homogeneity\footnote{Here statistical homogeneity
is understood as a scale at which the variance in density from one dust cell to
another is bounded, rather than a scale at which a FLRW model is approached.
The fact that this coincides with the BAO scale is a natural consequence of
initial conditions on the density perturbation spectrum at last scattering.
On scales smaller than the BAO scale the initial perturbations were slightly
amplified by acoustic oscillations in the primordial plasma, leading to a
greater variance in density contrasts on scales below the BAO scale as
compared with scales larger than the BAO scale. A crude estimate \cite{obs}
of the variance in density on scales larger than the BAO averaging scale gives
a variance of order 6\% in density on scales $\gsim100h^{-1}\,$Mpc, which
accords well with observations of order 7--8\% from galaxy clustering
statistics \cite{Hogg,SVBL}.}. It is hypothesized that within such a cell there
is a notion of average homogeneous expansion when CIFs in regions of different
density, which have undergone different relative volume decelerations, are
patched together appropriately. The proper volume of void regions increases
more rapidly, but there is a compensating increase of the clock rate of
isotropic observers\footnote{Isotropic observers are those who to leading
order see an isotropic CMB. Unlike the standard cosmology, however, in the
timescape scenario the mean CMB temperature will differ for ideal observers
within surfaces of average spatial homogeneity who have undergone varying
amounts of regional volume deceleration and consequently have differently
normalized clocks.} within voids, as compared to isotropic observers in the
denser wall regions. In this way there is always a choice of rulers and clocks
for which an average uniform expansion is maintained.

The dust ``particles'' are then regions of the cosmic fluid which contain great
variations in density and spatial curvature and which can be described by
different alternative choices of time and space coordinates in a smoothed out
description.  The different coordinate systems are those adapted to isotropic
observers in different locations in the fluid element, who have undergone
different relative amounts of regional volume deceleration, and who extend
coordinates to the whole element with a time coordinate assumed synchronous
to their own. By any one set of clocks it appears that the void regions
expand faster than the wall regions, and thus observers working with a
single clock will assign a variance to the expansion rate within each
fluid cell, even though there is another gauge in which the expansion is
uniform.

It is the equivalence of the different descriptions of the coarse-grained
fluid cell by isotropic observers in regions of different local Ricci scalar
curvature that replaces diffeomorphism invariance of the microscopic metric
as a relevant ``coordinate freedom'' of the coarse-grained metric description.
The Buchert time parameter is assumed to apply to those isotropic observers
whose locally measured spatial curvature is the same as the volume average
spatial curvature when averaged on horizon scales. The Buchert formalism
is assumed to apply insofar as the Buchert time parameter is a collective
coordinate of the coarse-grained fluid element, and the variance in expansion
rate refers to that attributed to the internal degrees of freedom of the fluid
element by an isotropic observer whose local spatial curvature scalar matches
the volume average one.

It is certainly true that the assumptions about use of the Buchert formalism
here represent an ansatz, which needs to be rigorously demonstrated in a
mathematical scheme for coarse-graining. However, since dust is not prescribed
in the Buchert formalism any attempt to use the Buchert formalism in
application to any realistic cosmology must invariably make assumptions about
how dust is to be defined. I take the view that one should begin by making
physically well-motivated assumptions, to see whether any phenomenologically
realistic model universe can be constructed. If that is the case, then an
appropriate mathematical formalism needs to be developed.

\subsection{Spatial averages versus null cone averages illustrated by the
timescape cosmology\label{obsd}}

If the Einstein equations (\ref{Efe}) are to be viewed as evolution equations
then a statistical description of average evolution would appear to have to
involve spatial averages, especially since the geometry at any event is
more influenced by the domain within the matter horizon \cite{ES09} than by
events on or close to the null cone. Nonetheless, almost all information
about cosmic evolution comes to us on our past null cone, and thus any attempt
to test a model of average cosmic evolution must relate the average
cosmological parameters to observations made on the past null
cone\footnote{For some other recent discussions of null cone averages
see \cite{Co09,GMNV}.}.

Even in the case of simple LTB models with prescribed dust, it is possible to
demonstrate that an average of cosmic expansion and acceleration on spatial
hypersurfaces does not in general coincide with the expansion and apparent
acceleration as measured on the past light cone \cite{BA}. The determination of
averages on the past light cone demands taking the position of the observer
into account. Consequently in any spatial averaging formalism, including the
Buchert formalism, specific arguments need to be presented for the
identification of cosmological parameters in terms of measurements on our
past light cone as observers in a galaxy.

Here I will briefly outline how these steps are achieved in the concrete
example of the timescape model \cite{clocks,sol,obs}. The timescape model
assumes that within dust cells coarse-grained at the $100h^{-1}\,$Mpc scale of
statistical homogeneity there are spatially
flat wall regions and negatively curved void regions. It is assumed we can
always enclose the bound structures which formed from over-critical density
perturbations within regions which are spatially flat on average, and
marginally expanding at the boundary. These boundaries are called
{\em finite infinity} regions \cite{fit1,clocks}, with local average metric
\beq\ds^2\Z{\Fi}=-c^2\dd\tw^2+\aw^2(\tw)\left[\dd\etw^2+\etw^2\dOM\right]\,.
\label{wgeom}\eeq
The walls constitute the disjoint union of such finite infinity regions.
\begin{figure}[htb]
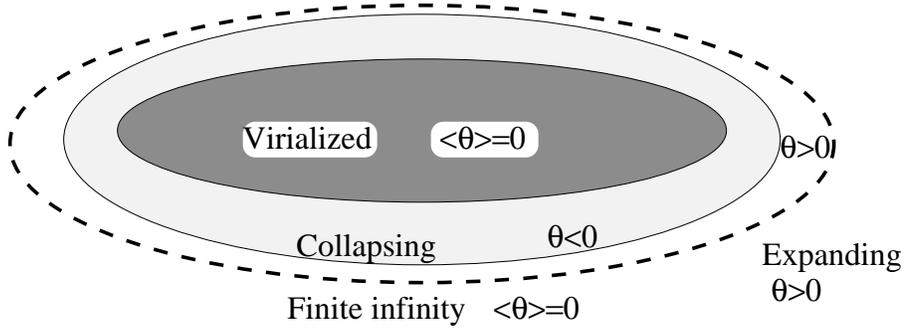

\vbox{\figfi
\caption{\label{fig_fi}%
{\sl A schematic illustration of the notion of finite infinity, $\hbox{\it fi}$
\cite{clocks}: the boundary (dashed line) to a region with average zero
expansion inside, and positive expansion outside. It may or may not contain
collapsing regions.}}}
\end{figure}

Different finite infinity regions will have different spatial extents,
being of order 1--2$\,$Mpc for our Local Group of galaxies, while being one
order of magnitude larger for rich clusters of galaxies. Although the overall
universe is inhomogeneous, finite infinity provides a demarcation between
bound systems and expanding regions (see figure~\ref{fig_fi}). It also provides
a notion of critical density regionally defined as the mass in a finite
infinity region divided by its volume. The boundaries of separate finite
infinity regions, although of different spatial extents, will have undergone
the same amount of volume deceleration since the epoch of last scattering
and the parameter $\tw$ is therefore assumed synchronous at all finite infinity
boundaries.

In addition to the walls there are the voids which dominate the present epoch
universe by volume. Voids will also have different spatial extents, being
characterized by regional negatively curved metrics of the form
\beq\ds^2\Z{\DD\ns{v}}=-c^2\dd\tv^2+\av^2(\tv)\left[\dd\etv^2+\sinh^2(\etv)
\dOM\right]\,.\label{vgeom}\eeq
Generally the voids will have different individual metrics (\ref{vgeom}).
However, in the centres of the $30h^{-1}$ Mpc voids \cite{HV1,HV2} the
regional metrics will rapidly approach that of an empty Milne universe
for which the parameters $\tv$ can be assumed to be synchronous.
One could potentially use different curvature scales for dominant voids
and minivoids to characterize the average scalar curvature $\ave{\cal R}$.
However, in the two--scale approximation of \cite{clocks,sol} a single
negative curvature scale is assumed as a simplification.

Although the construction is reminiscent of the Swiss cheese model, the
important difference is that the metrics (\ref{wgeom}) and (\ref{vgeom})
both represent the regional geometries in disjoint regions of different
spatial extents. Since there is no global FLRW metric there is no ``cheese''
in this construction. The global spacetime structure is determined instead
by a Buchert average.

The Buchert average is constructed as a disjoint union of wall and void
regions over the entire present horizon volume
$\Vav=\Vav\ns i\ab^3$, where
\beq\ab^3=\fvi\av^3+\fwi\aw^3,\label{bav}\eeq
$\fvi$ and $\fwi=1-\fvi$
being the respective initial void and wall volume fractions at last
scattering. It is convenient to rewrite (\ref{bav}) as $\fv(t)+\fw(t)=1$,
where $\fw(t)=\fwi\,\aw^3/\ab^3$ is the {\em wall volume fraction} and
$\fv(t)=\fvi\,\av^3/\ab^3$ is the {\em void volume fraction}. Within the
dust particles the metrics (\ref{wgeom}) and (\ref{vgeom}) are assumed to be
patched together with the condition of uniform quasilocal bare Hubble flow
\cite{equiv1,clocks} discussed in section~\ref{Machh},
\beq
\bH={1\over\aw}\Deriv\dd\tw\aw={1\over\av}\Deriv\dd\tv\av.
\eeq
Since this bare Hubble parameter is uniform within a dust particle, it will
also be equal to the Buchert average parameter Hubble parameter (\ref{Hav}).
For the purpose of the Buchert average it is convenient to refer all quantities
to the set of volume--average clocks that keep the time parameter $t$ of
(\ref{buche1})--(\ref{backr}) so that
\beq
\bH={\dot\ab\over\ab}=\gw\Hw=\gv\Hv
\eeq
where
\beq\Hw\equiv{1\over\aw}\;\Deriv\dd t\aw,\qquad
\Hv\equiv{1\over\av}\;\Deriv\dd t\av\,,\label{homo3}\eeq
and
\beq\gw\equiv\Deriv\dd\tw{t\ },\qquad\gv=\Deriv\dd\tv{t\ },
\label{clocks1}
\eeq
are phenomenological lapse functions of volume--average time, $t$, relative
to the time parameters of isotropic wall and void--centre observers
respectively. The ratio of the relative Hubble rates $h_r=\Hw/\Hv<1$ is related
to the wall lapse function by
\beq\gw=1+{(1-h_r)f_v\over h_r},\label{clocks2}\eeq
and $\gv=h_r\gw$.

In this two-scale approximation the Buchert equations for pressureless dust
with volume--average density $\rhb\Z M$ can be solved analytically \cite{sol}
if we make the assumption that volume--average shear is negligible and that
there is no backreaction within walls and voids separately\footnote{In the
model of Wiegand and Buchert \cite{WB}, two components of overdense and
underdense regions are similarly identified, but with two additional parameters
representing internal backreaction within these regions. Furthermore,
as well as having a different observational interpretation of their solutions,
Wiegand and Buchert's choice of the values of the parameters equivalent to the
initial fractions $\fwi$ and $\fvi$ is different, as they do not formally
identify walls and voids \cite{WB}. Since walls and voids do not exist at the
surface of last scattering, I assume that the vast bulk of the present horizon
volume that averages to critical density gives $\fwi\simeq1$, while $\fvi=1-
\fwi$ is the small positive fraction of the present horizon volume that
consists of {\em uncompensated} underdense regions at the last scattering
surface.}, but only in the combined average. With this assumption, the
kinematic backreaction term becomes \cite{clocks}
\beq\QQ=6\fv(1-\fv)\left(\Hv-\Hw\right)^2=
{2\dot\fv^2\over3\fv(1-\fv)}\,.\label{Q1}
\eeq
The independent Buchert equations (\ref{buche1}), (\ref{intQ}) then reduce to
two coupled nonlinear ordinary differential equations
\cite{clocks} for
$\ab(t)$ and $\fv(t)$, namely
\bea
&&\OMM+\OMk+\OMQ=1,\label{Beq1}\\
&&\ab^{-6}\pt_t\left(\OMQ\bH^2\ab^6\right)+\ab^{-2}\pt_t\left(\OMk\bH^2\ab^2
\right)=0\,,\label{Beq2}
\eea
where
\bea\OMM&=&{8\pi G\rhb\Z{M0}\ab\Z0^3\over 3\bH^2\ab^3}\,,\label{omm}\\
\OMk&=&{-\kv c^2\fvi^{2/3}\fv^{1/3}\over \ab^2\bH^2}\,,\label{omk}\\
\OMQ&=&{-\dot\fv^2\over 9\fv(1-\fv)\bH^2}\,,\label{omq}
\eea
are the volume--average or ``bare'' matter density, curvature density and
kinematic backreaction density parameters respectively, $\ab\Z0$
and $\rhb\Z{M0}$ being the present epoch values of $\ab$ and $\rhb\Z M$.
The average curvature is due to the voids only, which are assumed
to have $\kv<0$. Equations (\ref{Beq1}), (\ref{Beq2}) are readily integrated
to yield an exact solution \cite{sol}, which also possesses a very simple
tracking limit at late times with $h_r\to2/3$.

The exact solution \cite{sol} of the Buchert equations is of course just a
statistical description which is not directly related to any physical metric
that has been specified thus far. Since all cosmological information is
obtained by a radial spherically symmetric null cone average, we retrofit a
spherically symmetric geometry relative to an isotropic observer who measures
volume-average time, and with a spatial volume scaling as $\ab^3(t)$,
\beq
\dd\mean s^2=-c^2\dd t^2+\ab^2(t)\,\dd\etb^2+\Aa(\etb,t)\,\dOM.
\label{avgeom}
\eeq
Here the area quantity, $\Aa(\etb,t)$, satisfies
$\int^{\etb\X{\cal H}}_0\dd\etb\, \Aa(\etb,t)=\ab^2(t)\Vav\ns{i}
(\etb\Z{\cal H})/(4\pi)$, $\etb\Z{\cal H}$ being the conformal distance to
the particle horizon relative to an observer at $\etb=0$. The metric
(\ref{avgeom}) is spherically symmetric by construction, but is not a
LTB solution since it is not an exact solution of Einstein's equations, but
rather of the Buchert average of the Einstein equations.

In terms of the wall time, $\tw$, of finite infinity observers in walls
the metric (\ref{avgeom}) is
\beq\dd\mean s^2=-\gw^2(\tw)\,c^2\dd\tw^2+\ab^2(\tw)\,\dd\etb^2+\Aa(\etb,\tw)\,
\dOM\,.\label{avgeom2}\eeq
This geometry, which has negative spatial curvature is not the
locally measured geometry at finite infinity, which is given instead
by (\ref{wgeom}). Since (\ref{wgeom}) is not a global geometry, we
match (\ref{wgeom}) to (\ref{avgeom2}) to obtain a {\it dressed} wall geometry.
The matching is achieved in two steps. First we conformally match radial null
geodesics of (\ref{wgeom}) and (\ref{avgeom2}), noting that null geodesics are
unaffected by an overall conformal scaling. This leads to a relation
\beq
\dd\etw={\fwi^{1/3}\dd\etb\over\gw\fvf^{1/3}}\label{etarel}
\eeq
along the geodesics. Second, we account for volume and area factors by taking
$\etw$ in (\ref{wgeom}) to be given by the integral of (\ref{etarel}).

The wall geometry (\ref{wgeom}), which may also be written
\beq \ds^2\Z{\Fi}=-c^2\dd\tw^2+{\fvf^{2/3}\ab^2\over\fwi^{2/3}}
\left[\dd\etw^2+\etw^2\dOM\right]\,,
\eeq
on account of (\ref{bav}), is a local geometry only valid in spatially flat
wall regions. We now use (\ref{etarel}) and its integral to extend this
metric beyond the wall regions to obtain the dressed global metric
\bea
\ds^2&=&-c^2\dd\tw^2+{\ab^2\over\gw^2}\,\dd\etb^2+
{\ab^2\fvf^{2/3}\over\fwi^{2/3}}\,\etw^2(\etb,\tw)\,\dOM\nonumber\\
&=&-c^2\dd\tw^2+a^2(\tw)\left[\dd\etb^2+\rw^2(\etb,\tw)\,\dOM\right]
\label{dgeom}\eea
where $a\equiv\gw^{-1}\ab$, and
\beq\rw\equiv\gw\fvf^{1/3}\fwi^{-1/3}\etw(\etb,\tw).\eeq
While (\ref{wgeom}) represents a local geometry only valid in spatially flat
wall regions, the dressed geometry (\ref{dgeom}) represents an average
effective geometry extended to the cosmological scales,
parametrized by the volume--average conformal time which satisfies
$\dd\etb=c\,\dd t/\ab=c\,\dd\tw/  a$. Since the geometry on cosmological scales
does not have constant Gaussian curvature the average metric (\ref{dgeom}),
like (\ref{avgeom}), is spherically symmetric but not homogeneous.

If as wall observers we try to fit a FLRW model synchronous with our clocks
that measure wall time, $\tw$, we are effectively fitting the dressed
geometry (\ref{dgeom}), which is effectively the closest thing there is to a
FLRW geometry adapted to the rulers and clocks of wall observers. The
cosmological parameters we infer from taking averages on scales much
large than the $100h^{-1}\,$Mpc scale of statistical homogeneity will not be
the bare parameters $\bH$, $\OMM$, $\OMk$, and $\OMQ$, but instead the
{\em dressed Hubble parameter}
\bea\Hh\equiv{1\over a}\Deriv\dd\tw a
={1\over\ab}\Deriv\dd\tw\ab-{1\over\gw}\Deriv\dd\tw\gw
=\gw\bH-\Deriv{\dd}t\gw\,,\label{42}
\eea
and the {\em dressed matter density parameter}
\beq \Omega\Z{M}=\gw^3\OMM\,.\eeq
There is similarly a dressed luminosity distance relation
\beq\dL=a\Z0(1+z)\rw,\eeq where $a\Z0=\gb\ns{w0}^{-1}\ab\Z0$, and the
{\em effective comoving distance} to a redshift $z$ is $D=a\Z0\rw$, where
\beq\rw=\gb\ns{w0}\fvf^{1/3}
\int_t^{t\X0}{c\,\dd t'\over\gw(t')(1-\fv(t'))^{1/3}\ab(t')}\,,
\label{eq:dL}\eeq
and $1+z\equiv a\Z0/a=(\ab\Z0\gw)/(\ab\,\gb\ns{w0})$.

It was demonstrated in \cite{clocks,sol} that for realistic initial conditions
at last scattering the dressed deceleration parameter is negative at late
epochs, even though the volume-average bare deceleration parameter is positive.
In particular, the general solution \cite{sol} possesses a tracking limit
which is reached to within 1\% by a redshift $z\goesas37$. For the tracker
solution \cite{sol} the phenomenological lapse function, $\gw(t)$, void
fraction $\fv(t)$, and bare Hubble parameter, $\bH(t)$, are related by
\bea
\gw&=&\frn32t\bH(t)\\ &=&1+\half\fv(t)\\
&=&{9\fvn\Hb t+2(1-\fvn)(2+\fvn)\over2\left[3\fvn\Hb t+(1-\fvn)(2+\fvn)
\right]}\,,\label{gam2}
\eea
where $\Hb$ and $\fvn$ are the present epoch values of $\bH(t)$ and $\fv(t)$.
From the bare Hubble parameter we can construct a bare deceleration parameter,
\beq
\bq\equiv{-\ddot\ab\over\bH^2\ab}=\half\OMM+2\OMQ={2\fvf^2\over(2+\fv)^2}\,.
\label{qbare}\eeq

For the tracker solution, the time parameter $\tw$ of wall observers is related
to the Buchert volume--average time parameter, $t$, by
\beq
\tw=\frn23t+{4\OmMn\over27\fvn\Hb}\ln\left(1+{9\fvn\Hb t
\over4\OmMn}\right)\,, \label{tsol}
\eeq
where $\OmMn=\frn12(1-\fvn)(2+\fvn)$ is the present epoch dressed matter
density. At early epochs, as $t\to 0$, $\tw\goesas t$, but at later epochs
the two parameters differ by an amount restricted to the range $\frn23t<\tw
<t$. The dressed Hubble parameter (\ref{42}) then satisfies
\beq
H={2\over3t}+{\fv(t)[4\fv(t)+1]\over6t}
=\bH(t)+{\fv(t)[4\fv(t)-1]\over6t}\,,
\eeq
and can be used to construct a dressed deceleration parameter,
\beq
q\equiv{-1\over H^2a^2}{\dd^2a\over\dd\tw^2}
={-\fvf(8\fv^3+39\fv^2-12\fv-8)\over\left(4+\fv+4\fv^2\right)^2}\,.
\label{qtrack}\eeq
At early epochs when the void fraction is small both deceleration parameters
take the value, $q\goesas\bq\goesas\half$, as would be expected for an
Einstein--de Sitter universe. However, whereas the bare deceleration parameter
(\ref{qbare}) is always positive, the dressed deceleration parameter changes
sign at a value of $\fv\simeq0.59$ corresponding to a zero of the cubic in
the numerator of (\ref{qtrack}). At very late epochs both deceleration
parameters approach the Milne value, but with opposite signs: $\bq\to0^+$,
while $q\to0^-$. For parameter values with a good fit to supernovae data
\cite{LNW,SW}, $\fvn\goesas0.76$, and apparent acceleration typically begins
at a redshift $z\goesas0.9$. Apparent acceleration is therefore a transitional
feature, which is largest in the transition period during which voids become
dominant. During this epoch the variance of local geometry in galaxies from
the volume average geometry grows large, leading to a variance in the relative
calibration of clocks and rulers. 

\begin{figure}[htb]
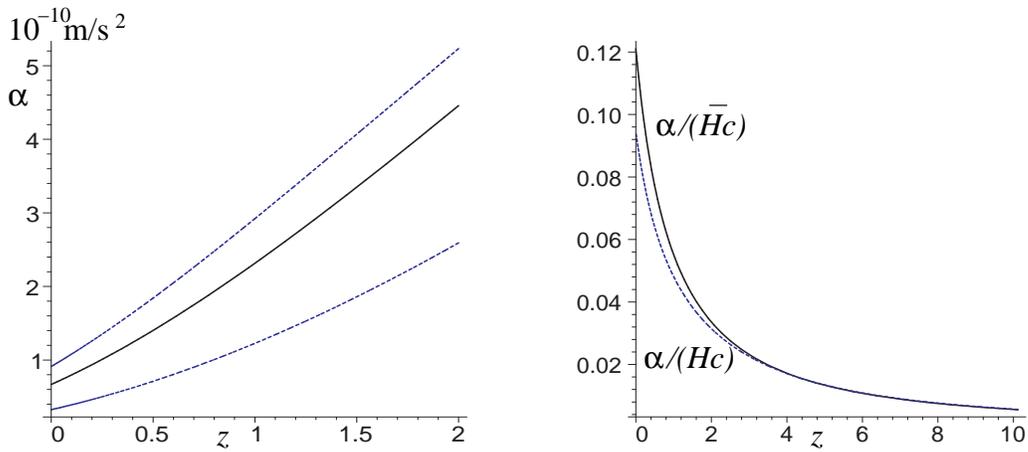

\vbox{\figaccel
\caption{\label{fig_a}%
{\sl The magnitude of the relative deceleration scale \cite{equiv1}, $\al$:
{\bf(a)} in terms of its absolute value for redshifts $z<0.25$; {\bf(b)} in
terms of the dimensionless ratios $\al/(c\bH)$ (solid curve) and $\al/(cH)$
(dashed curve) for redshifts $z<10$. In panel (b) just the best fit value
$\fvn=0.76$ is shown, whereas in panel (a) the solid and dashed represent the
best fit value and 1$\si$ uncertainties, $\fvn=0.76^{+0.12}_{-0.09}$
\cite{LNW} respectively. A value $\Hm=61.7\kmsMpc$ is assumed.}}}
\end{figure}

For parameter values which phenomenologically fit the observed expansion
history of the universe \cite{obs,LNW,SW}, we find $\gb\ns{w0}\goesas1.37$,
meaning that the variance in the two time parameters $\tw$ and $t$ in
(\ref{tsol}) typically grows to 37\% by the present epoch
\cite{clocks}. Such a large difference is counter-intuitive, but can be
understood as the cumulative integrated effect resulting from a tiny relative
deceleration of whose magnitude is defined in terms of the phenomenological
lapse function by \cite{equiv1}
\beq
{\al\over c}={1\over\sqrt{\gw^2-1}}\Deriv\dd\tw\gw=\Der\dd t
\sqrt{\gw^2-1}\,.
\label{a2}\eeq
Substituting the tracker solution (\ref{gam2}) in (\ref{a2}) gives a relative
deceleration as plotted in figure~\ref{fig_a}. Its value at the
present epoch is $\al\Z0\goesas7\times10^{-11}$m$\,$s$^{-2}$ and is typically
of order $10^{-10}$m$\,$s$^{-2}$ for most of the life of the universe.
Although the absolute value of $\al$ was higher in the past, the relative
expansion rate was even higher in the past. As a fraction of the Hubble
parameter, $\al/(Hc)$, is highly suppressed at early epochs: this
dimensionless ratio ranges from $0.1$ at the present
epoch to $6\times10^{-6}$ at the epoch of last scattering. Thus even though
we are dealing with an effect whose instantaneous magnitude is extremely small
and well within the weak field regime, it can still lead to a significant {\em
cumulative} difference when one has the age of the universe to integrate over.

\subsection{Observational tests of the timescape cosmology}

A variety of inhomogeneous model universes, including the LTB model and
other exact solutions \cite{BCK}, can be fit to observational data which
measure the expansion history of the universe in a variety of ways, with
varying degrees of success. One must take care with such fitting, however,
since the raw data has often been reduced assuming the standard homogeneous
cosmology in ways which can sometimes be very subtle.
Type Ia supernovae (SneIa), for example, are not standard candles but rather
are standardizable candles, and one widely used light curve fitter
\cite{Guy05,Guy07} marginalizes over parameters of the standard cosmology
as well as empirical light curves parameters when reducing the data.
Na\"{\i}vely using the reduced data in studies of inhomogeneous models
and averaging, which a number of researchers unfortunately do, can therefore be
problematic.

At the very least one needs a clear idea about how inhomogeneity limits
the derivation of an average expansion. In the timescape scenario, the
minimum scale on which an average isotropic Hubble law is expected is
the $100h^{-1}$Mpc scale of statistical homogeneity. Since many SneIa
data sets contain significant numbers of points below this scale, care
must be taken to remove such points, or more generally to consider
how the inclusion of such events might affect the overall calibration of
light curves.

Observational tests of the timescape cosmology are relatively well developed
\cite{obs,LNW,SW}, and to the extent that it has been tested the timescape
model is competitive with the standard \LCDM\ cosmology. Luminosity distances
of SneIa \cite{LNW,SW} are the most well tested. It
was recently shown \cite{SW} that in terms of current data sets the differences
between the luminosity distance predictions of the \LCDM\ and timescape
cosmologies are at the same level as systematic uncertainties in the light
curve fitters, due to unknown properties of SneIa. These include, in
particular, a degeneracy between the effect of intrinsic colour variations in
SneIa events and the effect of absorption by dust in the host galaxies,
which is currently being investigated by astronomers. Depending on which
light curve fitting method one uses, one can find that there is Bayesian
statistical evidence in favour of the standard \LCDM\ cosmology over the
timescape cosmology, or alternatively in favour of the timescape cosmology
over the \LCDM\ cosmology. In other words, there are already enough SneIa
events to distinguish the two models, but the empirical treatment of SneIa
light curves to convert them to standard candles still needs to be
understood before conclusions can be drawn.

Our recent study of SneIa luminosity distances \cite{SW} finds that
making cuts to the data below the $100h^{-1}$Mpc scale of statistical
homogeneity can significantly affect the analysis. Furthermore, in terms of
SneIa systematics, we find that the timescape scenario would appear to
be more obviously favoured over the \LCDM\ model if the reddening
law for dust in other galaxies has a reddening parameter, $R\Z V$, close
to the value $R\Z V\simeq3.1$ observed for dust in the Milky Way \cite{SW},
rather than half this value. Since the reddening law in nearby galaxies
can be tested independently, future investigations of such issues will have the
power to falsify or strongly constrain the timescape scenario. Thus far
such studies find values $R\Z V=2.82\pm0.38$ and $R\Z V=2.71\pm0.43$
for two samples of eight and seven nearby galaxies respectively
\cite{Fin08,Fin11}, consistent with the Milky Way value.

In the timescape model parameter values have also been determined which fit the
angular diameter distance of the sound horizon in the CMB anisotropy spectrum,
and the BAO scale in galaxy clustering statistics \cite{obs,LNW,SW}. These
estimates are crude ones at this stage, as the raw data still needs to be
reanalysed from first principles using the timescape methodology before
statistical bounds can be obtained, and this is likely to be a very involved
process. Several potential tests of the expansion history were proposed in
\cite{obs}, including, for example, variants of the Alcock-Paczy\'nski test
\cite{AP}, the inhomogeneity test of Clarkson, Bassett and Lu \cite{CBL},
and the time drift of cosmological redshifts \cite{zd1,zd2,zd3}. Furthermore,
three separate tests with indications of results in possible tension with the
\LCDM\ model on the basis of existing data are found to be consistent with the
expectations of the timescape cosmology \cite{obs}. Since these observational
tests have already been briefly reviewed elsewhere \cite{obsr}, I will not
discuss them further here.

The greatest observational challenge for the timescape model is the value of
dressed Hubble constant on scales larger than that of statistical homogeneity.
If we compare the angular diameter distance of the sound horizon seen in the
CMB anisotropy spectrum and the effective comoving scale of the BAO as seen
in galaxy clustering statistics, then a range of values of the dressed Hubble
constant, $57\lsim\Hm\lsim68\kmsMpc$, would be admissible in the timescape
scenario \cite{SW}. This is at odds with the recent measurement of
$\Hm=73.8\pm2.4\kmsMpc$ by the SHOES survey \cite{R11}. However, it is a
feature of the timescape model that a 17--22\% variance in the apparent Hubble
flow will exist on local scales below the scale of statistical homogeneity,
and this may potentially complicate calibration of the cosmic distance
ladder.

Further quantification of the variance in the apparent Hubble flow in
relationship to local cosmic structures would provide an interesting
possibility for tests of the timescape cosmology for which there are no
counterparts in the standard cosmology. There is evidence from the study of
large-scale bulk flows that apparent peculiar velocities determined in the
FLRW framework have a magnitude arguably in excess of the statistical
expectations of the standard \LCDM\ model \cite{WFH,K09,K10}. In the timescape
model it conceptually better to think in terms of varying expansion
rates, rather than peculiar velocities. Nonetheless, given that our location
is right on an edge between a wall and a dominant void \cite{Tully} the
effective equivalent maximum peculiar velocity can be estimated as
\beq
v\ns{pec}=(\frn32\Hb-\Hm){30\over h}\w{Mpc}=510^{+210}_{-260}\w{km/s}\,.
\eeq
This estimate assumes the typical diameter of $30\,h^{-1}\,$Mpc for the local
void, and uses the tracker solution relation $\Hb=2(2+\fvn)\Hm/\left(4\fvn^2+
\fvn+4\right)$ \cite{sol} with the best fit values $\fvn=0.76^{+0.12}_{-0.09}$
\cite{LNW}. This rough estimate is of a magnitude consistent with observation.

In any inhomogeneous cosmology the manner in which we estimate peculiar
velocities from the data needs to be carefully considered. In the case of the
kinematic Sunyaev-Zel'dovich effect \cite{K09,K10} the standard FLRW model is
assumed in the data reduction in possibly subtle ways. One important question
is whether the CMB dipole is purely due to our peculiar velocity with respect
to the surface of average homogeneity, or whether it also contains some
fraction, perhaps just at the percent level, which is due to foreground
inhomogeneities within the scale of statistical homogeneity.

A more insidious problem with conceptual thinking is that many researchers,
particularly observationalists, tend to think in terms of a uniform FLRW
expansion of the universe in Euclidean space with an action--at--a--distance
Newtonian gravitational force embedded on top, so that we experience
``infall'' towards the Virgo cluster or even towards the much more distant
Shapeley concentration even though the physical distances to these objects are
always increasing. There is no fundamental reason to expect spacetime to
arrange itself so that the intuition we have about gravity from the solar
system repeats itself on the very largest of scales. If we use just one set of
clocks, then in any inhomogeneous model it makes more sense conceptually
to think about variations of the expansion rate in regions of different
density (and spatial curvature) which decelerate by different amounts.
We need to think a bit more deeply in analysing the data, particularly given
the conundrum that observed peculiar velocities of galaxies with respect to a
FLRW background do not match statistical expectations.

\section{Discussion}

In this review I have discussed what I believe are some of the most important
physical questions in relation to the averaging problem. In my view, we stand
at a very exciting juncture for the development of general relativity, as
the mystery of dark energy indicates that deep fundamental questions remain
concerning our understanding of spacetime on the largest of scales. Even if
dark energy is ``just'' a cosmological constant, then that would be of
profound significance, since quite apart from the problem of explaining its
magnitude we would have to understand why there is a field which permeates
spacetime without reference to other matter.

I believe that it is time to more seriously consider Einstein's dictum that
{\em``In a consistent theory of relativity there can be no inertia relatively
to `space', but only an inertia of masses relatively to one another''}
\cite{esu}. In particular, rather than modifying gravity to add exotic fields
in the vacuum in ways which potentially violate the weak equivalence principle,
we should consider modifications that do not violate any existing principles
but which might add limiting principles to give a deeper realisation of
spacetime as a relational structure, consistent with Mach's principle. The
cosmological equivalence principle (CEP) \cite{equiv1,equiv2} is proposed with
such an end in mind. Although there may ultimately be better ways of framing
relevant principles, one cannot escape from the fact that if the averaging
problem is to be thought of in physical terms, then it is intimately related to
the cosmological statement of Mach's principle \cite{Bondi}.

The observed universe has a very complex hierarchical structure \cite{SL11},
and is very clearly inhomogeneous on scales $\lsim100h^{-1}\,$Mpc. This has led
many researchers to consider both exact inhomogeneous solutions of general
relativity \cite{BCK}, as well as the averaging problem for
general inhomogeneous metrics. Yet the vast majority of this effort is
mathematically driven, rather than physically driven.

Why is the expansion so close to that of a FLRW model despite the observed
inhomogeneity? A potential answer is that there is a canonical choice of clocks
and rulers that can always be made in the averaging problem to make the
regional expansion uniform -- analogously to the freedom of choosing Riemann
normal coordinates to make the first derivatives of the metric zero near a
point -- and it is this choice which is made by nature to define average
spatial homogeneity and preserve the near isotropy of the CMB.

Without additional limiting principles inhomogeneous geometries offer
so many potential parameters that it is difficult to see how they could be
constrained. Many researchers choose to limit their models by the demand that
the average evolution is a FLRW one, as discussed in section~\ref{FLRWa}. Yet
the FLRW universe is not singled out by any physical principle, and it
embodies three separate notions of average spatial homogeneity which may well
be overly restrictive.

I suggest that the CEP, or something close to it, is a limiting principle which
singles out a notion of uniform expansion as the condition of average spatial
homogeneity without necessarily leading to FLRW evolution.
A CMC-like slicing generalizes the notion of relating inertial frames by a
uniform velocity, and has been independently recognised by a number of
researchers \cite{BKL07,equiv1,ABFKO} as embodying Mach's principle.

The timescape scenario is a framework which attempts to put such physical
principles into a simple cosmological model. As a phenomenological model it is
interesting to note that it is competitive with the standard \LCDM\ model, in
as far as it has been tested to date \cite{obs,SW,obsr}, with the only obvious
major challenge at present being the value of the global average Hubble
constant \cite{R11}. However, there are many outstanding issues in the
timescape scenario, which prevent many researchers from giving it further
consideration.

One clear problem is the issue of junction conditions and the patching
together of CIFs to realise the uniform Hubble flow condition within a dust
``particle''. In the two scale model outlined in section~\ref{obsd} the wall
and void regions are combined in a disjoint union without applying junction
conditions\footnote{The effect of junction conditions can be seen in the case
of LTB models with prescribed dust or networks of such LTB voids. In these
cases the shear in hypersurfaces of constant comoving time counteracts the
variance in volume expansion leading to a greatly suppressed backreaction
\cite{MM1,MM2}. The timescape scenario deals with a rather different situation:
in particular, it does not deal with prescribed dust constrained to avoid shell
cross singularities, nor with highly symmetric exact solutions which have been
cut and pasted together. Finally, the surfaces of average spatial homogeneity
are not assumed to be surfaces of constant comoving time.
In the LTB void model a ``uniform Hubble flow'' slicing of the prescribed dust
for reasonable density contrasts can only be introduced at the price of
taking hypersurfaces which are not necessarily purely spacelike \cite{Mn}.}.
The reason this has not yet been done is that it would require the development
of mathematical tools for the coarse-graining of geometries in a statistical
sense, and this is far from trivial. It is not a simple case of cutting and
pasting exact solutions for prescribed dust by well-known
techniques. Rather the solution of the problem is intimately related to the
question of what a dust fluid element is in general relativity when we have
to coarse grain over gravitational degrees of freedom themselves.

One feature of the formalism that is required to tackle this problem
is that it should deal with regional symmetries.
In particular, whereas general relativity deals with
diffeomorphism invariance on one hand, and the point symmetries of the
Lorentz group on the other, establishing CIFs requires us to deal with
the collective degree of freedom corresponding
to a regional volume expansion in particular. Different CIFs which have
undergone different amounts of relative volume deceleration will have
differing phenomenological lapse functions.

The construction of a phenomenological lapse function\footnote{If the only
symmetries that are allowed are diffeomorphisms of the global metric on one
hand, and local Lorentz transformations corresponding to rotations and boosts
on the other, then realistically there is no room for clock rate variations
of the order of magnitude dealt with in the timescape scenario \cite{R10}.
However, the suggestion here is that additional mathematical ingredients are
required to define regional symmetries when coarse-graining.} from a more
rigorous mathematical basis requires careful thought. In particular, there
is no single global ADM metric covering the whole universe which adequately
describes the metric degrees of freedom associated with galaxy clusters.
Thus it is not a simple matter of taking a single ADM lapse function and
integrating it out when coarse-graining. Furthermore, since we
are dealing with a collective degree of freedom corresponding to a regional
volume expansion, the phenomenological lapse function is also subtly different
to the gamma factor in a Lorentz boost. By the semi-tethered lattice analogy
\cite{equiv1,equiv2} it is not associated with a boost in any particular
direction but has more the character of a boost which is orthogonal to every
spatial direction. Since a rigorous treatment requires a new as
yet undeveloped methodology, it is difficult to convince sceptics of its
necessity. However, from a physical standpoint a relative deceleration implies
a difference in the amount of kinetic energy of expansion that is converted to
other forms of energy through gravitational collapse, which must
have very tangible consequences for physical processes. To ignore this problem
-- simply because it involves the thorny issue of the nature of gravitational
energy -- is to ignore some of the most fundamental principles of physics.

Dealing with the regional average symmetries that emerge in coarse-graining
inevitably means that we must consider quasilocal quantities, and in
particular quasilocal mass--energy and angular momentum. On account of the
strong equivalence principle we can always get rid of gravity near a point, and
so the definition of quasilocal energy is a subtle problem, which has been
studied for decades without any clear consensus emerging. (For a review see
\cite{quasi_rev}.) The quasilocal energy problem has principally been studied
for isolated systems, where conventional notions of mass associated with
asymptotically flat systems are well grounded. The issue of quasilocal
energy is relatively little studied for cosmological solutions, and
what work there is usually makes reference to specific exact solutions such
as the FLRW models \cite{CLN}--\cite{WCLN}. Sussman has considered the
specific case of defining relevant quasilocal variables for generalized LTB
models \cite{S08a} and their relationship to averaging and backreaction for
the case of spherically symmetric dust \cite{S08b,S08c,S10}.

Rather than always working with the same set of exact cosmological solutions,
more effort is needed to understand quasilocal variables that might be relevant
for more general coarse-graining procedures. Korzy\'nski's approach \cite{Ko09}
represents an interesting idea, which still remains to be fully developed.
The statistical nature of gravitational energy and entropy on the largest
scales is an unsolved fundamental problem which might be better understood
by thinking more carefully about these procedures.

In summary, it is my view that the apparently accelerated expansion of the
universe demands that we take a fresh look at the foundations of cosmological
general relativity from first principles. In particular, we face the very
real possibility that ``dark energy'' is simply an illusion due to our
misunderstanding of gravitational energy gradients in a complex hierarchical
geometry. To attack a problem as fundamental as gravitational energy we
must think fundamentally.

The argument about whether it is
better to use spacetime averages (section~\ref{secZ}) as opposed to spatial
averages (section~\ref{secB}) cannot really be addressed without asking the
more basic question of what is the structure of spacetime on the largest
of scales, especially over scales larger than that of the matter horizon
\cite{ES09} beyond which the exchange of particles and energy between
observers is minimal. Rather than simply taking a principle such as general
covariance as being paramount, we have to ask why was general covariance
introduced? The reason was that it is a way of characterizing physical
laws which combine gravity with the nongravitational interactions
of nature in such a way that spacetime geometry is a relational structure
between the elementary particles which interact via nongravitational forces.

In seeking to coarse grain gravitational degrees of freedom themselves, we
have to be prepared for the possibility that realising spacetime as a
relational structure might involve new ingredients beyond those which apply
to nongravitational microphysics or general relativity on the scale of
isolated systems. For example, the Bianchi I universe picks
out preferred directions in space, and is at odds with observation; but would
not be admitted by the CEP. Whether the CEP or some other principle is the
correct one, what is most important is that we take up the challenges offered
by cosmological observations to think more deeply about the foundations of
general relativity as a physical theory of the universe.

\ack

This work was supported by the Marsden Fund of the Royal Society of New
Zealand. I thank Teppo Mattsson for helpful discussions. This paper was
written in a home recovering from the 2011 Christchurch earthquake -- I wish
to thank my family for their forbearance.

\end{document}